\documentclass[english,prd,onecolumn,nofootinbib]{revtex4}
\usepackage[final]{graphicx}
\usepackage{bm, color}
\usepackage[colorlinks=false]{hyperref}
\usepackage{amsmath,mathrsfs,bbm}
\usepackage{wrapfig}
\usepackage[leftcaption]{sidecap}
\usepackage{color}

\def\be{\begin{equation}}
\def\ee{\end{equation}}
\def\bea{\begin{eqnarray}}
\def\eea{\end{eqnarray}}

\def\d {\mathrm{d}}
\newcommand{\HH}{\mathcal{H}}

\newcommand{\p}{_{{\text{\tiny$\|$}}}}

\newcommand{\e}{{\bm{{p}}}}
\newcommand{\pe}{{{{p}}}}

\newcommand{\J}{{\bm{\mathcal{J}}}}
\newcommand{\K}{{\bm{\mathcal{K}}}}
\newcommand{\I}{{\bm I}}
\newcommand{\C}[3]{{{\mathscr{C}}^{#2(#1)}_{#3}}}
\renewcommand{\S}[3]{{\mathscr{S}}^{#2(#1)}_{#3}}
\newcommand{\F}{{\bm F}}
\newcommand{\x}{{\bm \xi}}

\newcommand{\R}{{\bm{\mathcal{R}}}}

\renewcommand{\i}{\mathrm{i}}

\begin{document}

\title{Roulettes: A weak lensing formalism for strong lensing  \\---  II. Derivation and analysis ---}
\author{Chris Clarkson}
\address{School of Physics \& Astronomy, Queen Mary University of London, Mile End Road, London E1 4NS, UK\\
and, Department of Mathematics \& Applied Mathematics, University of Cape Town, Cape Town 7701, South Africa.
}
\email{chris.clarkson@qmul.ac.uk}
\date{\today}

\date{\today}

\begin{abstract}

We present a new extension of the weak lensing formalism capable of describing strongly lensed images. This paper accompanies \href{http://arxiv.org/abs/1603.04698}{Paper I}, where we provide a condensed overview of the approach and illustrated how it works. Here we give all the necessary details, together with some more explicit examples.  We solve the non-linear geodesic deviation equation order-by-order, keeping the leading derivatives of the optical tidal matrix, giving rise to a series of maps from which a complete strongly lensed image is formed. The family of maps are decomposed by separating the trace and trace-free parts of each map. Each trace-free tensor represents an independent spin mode, which distort circles into a variety of roulettes in the screen-space. It is shown how summing this series expansion allows us to create large strongly lensed images in regions where convergence, shear and flexion are not sufficient.

\end{abstract}

\maketitle

This paper is a detailed exposition of Paper I, \href{http://arxiv.org/abs/1603.04698}{arXiv:1603.04698},~\cite{Clarkson:2016zzi}, which presents the key elements of the subject matter in a wider context.

\tableofcontents

\section{Introduction}

In cosmology the weak lensing formalism is usually used for making lensing maps, which then give the dark matter distribution. The formalism assumes an image has distortion from a shear and convergence, and that these are constant over the image.  For extended sources or more strongly lensed objects this may not be appropriate as the shear and convergence can vary appreciably over an image. The first-order correction to this has been calculated and termed flexion, and is shown to give some arc-like features to images~\cite{Goldberg:2004hh,Bacon:2005qr,Clarkson:2015pia}. Rather surprisingly, the extension of this idea to arbitrary lens strength has not been given. 

We show here how to extend this idea to arbitrary order. The idea is to build up a family of higher-order maps, generalising the linear Jacobi, and second-order Hessian maps~\cite{Clarkson:2015pia}, from which a complicated image can be formed by summing over these (constant) maps.
The physical origin of this comes from the non-linear geodesic deviation equations, which describes how non-infinitesimally separated geodesics deviate from each other~\cite{Vines:2014oba}. This is a phenomenally complicated equation, but for lensing only a few terms are important. These are the ones with the highest number of derivatives in the screen space, for which the calculation becomes relatively straightforward. By integrating this in a perturbative way, we can construct higher-order lensing maps and extensions of the amplification matrix of normal weak lensing. At each order an amplification tensor can be decomposed into its various trace free parts which form the invariant parts~-- spin components~-- of the map, analogous to the convergence (trace) and shear (trace-free) as invariants of the amplification matrix. Each invariant mode (a `roulette') distorts an image in a peculiar way, such that the sum over all of them is able to reconstruct a complete image. We show also that an image can be inverted to give the amplitudes (in principle) of each mode, giving a  correspondence between a lens and strongly distorted image. In a sense, then, this expansion is the natural expansion for lensed images. 

In the following section, we formulate the non-linear geodesic deviation equation in the way we need it, and integrate it to give the maps at each order. We then show how we can decompose this map into its independent trace-free parts which are the spin modes of each map. We then discuss the formalism in the weak field approximation, and give a couple of simple examples of the roulette series in action. We then conclude, and give an appendix on some fundamental trig integrals which appear in the roulette amplitudes. 

\section{The non-linear geodesic deviation equation and the lensing maps}

The computation of the convergence and shear is achieved from the geodesic deviation equation which is linear in the deviation vector~$\xi^a$ which joins two neighbouring null geodesics of a congruence $k^a$:\footnote{We follow the notation of~\cite{Clarkson:2015pia}. $a,b,c,\cdots$ denote spacetime indices,  $k$ or $\xi$ as an index denotes projection of that index in the direction of $k$ or $\xi$. $A,B,C,\cdots$ are tetrad indices in the screen space using the Sachs basis $e_A^{~~b}$, and those indices are raised and lowered with $\delta_{AB}$. $R^a_{~bcd}$ is the Riemann tensor. A dot is a derivative along the null curve $=k^a\nabla_a$. We use $\lambda$ as the affine parameter of $k^a$ and $\lambda_o$ ($\lambda_s$) is its value at the observer (source). We will use bold symbols to denote vectors and matrices (rank-2 tensors) in the Sachs basis, and use standard matrix notation for these. For higher-order tensors in the Sachs basis we will keep using index notation.}
\be\label{sdjbcsbds}
\ddot \xi^a + R^a_{~k b k}\xi^b=0\,.
\ee
This applies for an infinitesimal deviation vector only as the Riemann tensor arrises from the first term in a series expansion over the vector.
The fully non-linear geodesic deviation equation is extremely complicated, given by the second derivative of Synge's world function, and its series expansion beyond second-order in $\xi^a$ is equally horrendous~--see \cite{Vines:2014oba}. As far as gravitational lensing is concerned, however, the most important contributions arise from terms with the maximum number of screen space derivatives of the metric (or rather the rotation coefficients). In this situation, we can `calculate' the leading contributions at each order in $\xi^a$, by recalling the derivation of \eqref{sdjbcsbds}. Considering two nearby geodesics, one Taylor expands the Christoffel symbols at the `end' of the vector $\xi^a$ about the point at the `start' of $\xi^a$. In this way the Riemann tensor appears as the linear part of this Taylor expansion of the Christoffel symbols in deriving \eqref{sdjbcsbds}:
\be
R^a_{~k \xi k} = \partial_\xi\Gamma^a_{~kk}-\partial_k\Gamma^a_{~k\xi}~~ (+ \Gamma\Gamma~\mbox{terms})\,.
\ee
Only the first term is important for us as all the others are sub-dominant. If in the standard textbook derivation of \eqref{sdjbcsbds} we keep more terms in the expansion leading to $\partial_\xi\Gamma^a_{~kk}$, we have $\sum_{n=1}\partial^{(n)}_\xi\Gamma^a_{~kk}/n!$, plus subdominant terms.
We can then replace the Riemann tensor in \eqref{sdjbcsbds} by 
\be
R^a_{~k b k}\mapsto\sum_{n=0}^\infty\frac{1}{(n+1)!}
(\xi^b\nabla_{b})^n R^a_{~k b k}\,.
\ee
The extra $n+1$ in the factorial comes about because the first term in the Riemann tensor is already the first term in the series expansion. 
Now, each $\xi^a\nabla_a$ can be replaced by $\xi^A\nabla_A$, as it is the screen space derivatives which are most important.
Now substitute this into~\eqref{sdjbcsbds}, and project into the screen space, and we have
\be\label{kdjscndnsck}
\ddot \xi^A - \mathcal{R}^A_{~~B}\xi^B=\sum_{n=1}^\infty\frac{1}{(n+1)!}
\xi^{A_1}\xi^{A_2}\cdots\xi^{A_n}\xi^B
\nabla_{A_1}\nabla_{A_2}\cdots\nabla_{A_n} \mathcal{R}^A_{~~B}\,,
\ee
where $\mathcal{R}_{AB}=-R_{AkBk}$ is the optical tidal tensor. Solutions to this may be found by an extension of the method used in~\cite{Clarkson:2015pia}. We can solve this perturbatively in powers of $\xi^A$, by writing
\be
\xi^A=\sum_{m=1}^{\infty}\frac{\epsilon^m}{m!}\xi^A_{(m)}\,,
\ee
where the power of $\epsilon$ is just to help us keep track of terms. Inserting this into \eqref{kdjscndnsck} we have
\be
\sum_{m=1}^{\infty}\frac{\epsilon^m}{m!}\left(\ddot \xi^A_{(m)} - \mathcal{R}^A_{~~B}\xi^B_{(m)}
\right)=
\sum_{n=1}^\infty\frac{1}{(n+1)!}\nabla_{A_1}\nabla_{A_2}\cdots\nabla_{A_n} \mathcal{R}^A_{~~A_0}\prod_{i=0}^n\sum_{m_i=1}^{\infty}\frac{\epsilon^{m_i}}{m_i!}\xi^{A_i}_{(m_i)}\,.
\ee
We solve this by equating powers of $\epsilon$ on both sides. For $\epsilon^m$, the term on the right with the largest number of derivatives of $R_{AB}$ has $m-1$ derivatives, and $m_i=1$ for each $i$. This implies that the $m$'th term in the solution obeys, to leading order,
\be\label{dsjkbcsjkdbcskcdb}
\ddot\xi^A_{(m)} - \mathcal{R}^A_{~~B}\xi^B_{(m)} =  \xi^{A_1}_{(1)}\xi^{A_2}_{(1)}\cdots\xi^{A_m}_{(1)}\nabla_{A_1}\nabla_{A_2}\cdots\nabla_{A_{m-1}} \mathcal{R}^A_{~~A_m} = F^A_{(m)}\,.
\ee
The solution to this equation is found by the method presented in \cite{Clarkson:2015pia}. There it was shown that the particular solution to 
\be\label{skdjncscn}
\ddot\x-\R\x=\F
\ee
is
\be\label{jdshbvshbdshbv}
\x=\int_{\lambda_o}^\lambda\d\lambda'\left[
\K(\lambda)-\J(\lambda)\J^{-1}(\lambda')\K(\lambda')
\right]\J^T(\lambda') \F(\lambda') \,,
\ee
where $\mathcal{J}_{AB}$ is the Jacobi map and $\mathcal{K}_{AB}$ is the reciprocal Jacobi map (written using standard matrix notation above for simplicity), which are linearly independent solutions to the linear GDE: 
\bea
\ddot\J=\R\J~~~&\text{with}&~~~\K(\lambda_o)=0,~~~\dot\J(\lambda_o)=-\I\,.
\\
\ddot\K=\R\K~~~&\text{with}&~~~\K(\lambda_o)=\I,~~~\dot\K(\lambda_o)=0\,.
\eea

\subsection{Initial and boundary conditions, and the lensing map at any order} There are two relevant boundary or initial conditions we can take to give useful solutions to \eqref{dsjkbcsjkdbcskcdb}: The image-to-source ($\searrow$) condition and the source-to-image ($\nwarrow$). Consider the image-to-source case first, where we give the initial conditions at the observer as
\be
(\searrow)~~~~~~~~~{\bm\xi}\big|_\text{observer}=0,~~~~\dot{\bm\xi}\big|_\text{observer}=-\bm\zeta\,.
\ee
Here $\bm\zeta$ is the non-perturbative angle in the image plane at the observer. At first order the solution is
\be\label{sjdncskdnj}
\xi_{(1)}^A=\mathcal{J}^A_{~~B}\zeta^B\,,
\ee
which gives the physical distance between two rays at the source to first order given the (exact) observed angular position between the two rays, $\bm\zeta$. The $m$'th order solution is~\eqref{jdshbvshbdshbv} plus a homogeneous solution of the same form as~\eqref{sjdncskdnj}. Since the initial conditions are satisfied for the first-order part, the homogeneous solutions at higher order must be zero. 
 At each order, then, lensing can be given by the $m$'th-order map, for $m\geq2$,
\be\label{dsjcsjdcn}
(\searrow)~~~~~~~~~~\xi_A^{(m)} = \mathcal{M}^\searrow_{AB_1\cdots B_m}\zeta^{B_1}\cdots\zeta^{B_m}\,,
\ee
where
\bea\label{djksncskjdnsjkcfvfkbjdfovj}
\mathcal{M}^{{}^\searrow}_{AB_1\cdots B_m}(\lambda) &=& 
\int_{\lambda_o}^\lambda\d\lambda'\big[\mathcal{K}_A^{~\,F}(\lambda)
-\mathcal{J}_A^{~\,D}(\lambda)(\mathcal{J}^{-1})_D^{~~E}(\lambda')
\mathcal{K}_E^{~\,F}(\lambda')
\big]\mathcal{J}_{~~F}^{G}(\lambda')\nonumber\\&&
\mathcal{J}^{~\,C_1}_{B_1}(\lambda')\cdots\mathcal{J}^{~\,C_m}_{B_m}(\lambda')
\nabla_{C_1}\cdots\nabla_{C_{m-1}} \mathcal{R}_{GC_m}(\lambda')\,.
\eea
This family of maps are written in the conventional way whereby $\bm\zeta$ is in the image plane (i.e., the \emph{angular} separation between two observed rays) is mapped to $\bm\xi$ in the source plane (the \emph{physical} distance between the same two rays at emission.) For $m=1$ we write ${\mathcal{J}}_{AB}=\mathcal{M}^{{}^\searrow}_{AB}$, noting that the Jacobi map is defined in this way. Adding up the $\bm\xi_{(m)}$'s for an observed deviation angle $\bm\zeta$ will give the physical position of the source element (assuming the maps are known). 

Although conventional, this is not necessarily  the most intuitive way round: we often want to predict how a source gets distorted.  
In this case we are interested in the source-to-image case with the boundary conditions
\be
(\nwarrow)~~~~~~~~~{\bm\xi}\big|_\text{observer}=0,~~~~{\bm\xi}\big|_\text{source}=\bm\eta\,,
\ee
where $\bm\eta$ is the (non-perturbative) physical distance between two rays at the source. The linear solution is the same as \eqref{sjdncskdnj} but now we write
\be
\bm\zeta_{(1)}=\J^{-1}(\lambda_s)\bm\eta\,,
\ee
to reflect the fact that the physical distance at the source  is given, and the observed (linear) deflection angle at the observer is derived. Consequently, the solution at any point on the geodesic, $\x_{(1)}(\lambda)$, with the same angular size at the image, which we require in \eqref{dsjkbcsjkdbcskcdb}, is given by
\be
\x_{(1)}(\lambda)=\J(\lambda)\J^{-1}(\lambda_s)\bm\eta\,.
\ee
The solution for $\x_{(m)}$ at order $m$ must be zero at the source since we have $\x_{(1)}(\lambda_s)=\bm\eta=\x(\lambda_s)$. Consequently,
\be
0=\x_{(m)}(\lambda_s)=\x^\text{homo}_{(m)}(\lambda_s)+\int_{\lambda_o}^{\lambda_s}\d\lambda'\left[
\K(\lambda)-\J(\lambda)\J^{-1}(\lambda')\K(\lambda')
\right]\J^T(\lambda') \F_{(m)}(\lambda') \,,
\ee
where
\be
\dot{\x}_{(m)}\big|_\text{observer}=\dot{\x}^\text{homo}_{(m)}\big|_\text{observer}\equiv-\bm\zeta_{(m)}\,,
\ee
and since we must have 
\be
\x^\text{homo}_{(m)}=\J\bm\zeta_{(m)}\,,
\ee
we find that the $m$'th perturbation to the angular size at the image is given by
\be
\bm\zeta_{(m)}=-\int_{\lambda_o}^{\lambda_s}\d\lambda'\left[
\J^{-1}(\lambda_s)\K(\lambda_s)-\J^{-1}(\lambda')\K(\lambda')
\right]\J^T(\lambda') \F_{(m)}(\lambda')\,.
\ee
The lensing maps in the source-to-image configuration are then
\be\label{kajsnckdjnaskjnd}
(\nwarrow)~~~~~~~~~~\zeta_A^{(m)} = \mathcal{M}^\nwarrow_{AB_1\cdots B_m}\eta^{B_1}\cdots\eta^{B_m}\,,
\ee
where
\bea\label{eorigjiorejgeirjogeoeijg}
\mathcal{M}^{{}^\nwarrow}_{AB_1\cdots B_m}(\lambda) &=& 
-\int_{\lambda_o}^\lambda\d\lambda'\big[(\mathcal{J}^{-1})_A^{~\,D}(\lambda)\mathcal{K}_D^{~\,F}(\lambda)
-(\mathcal{J}^{-1})_A^{~\,D}(\lambda')
\mathcal{K}_D^{~\,F}(\lambda')
\big]\mathcal{J}_{~~F}^{G}(\lambda')\nonumber\\&&
\mathcal{J}^{~\,C_1}_{B_1}(\lambda')\cdots\mathcal{J}^{~\,C_m}_{B_m}(\lambda')
(\mathcal{J}^{-1})^{~\,D_1}_{C_1}(\lambda)\cdots(\mathcal{J}^{-1})^{~\,D_m}_{C_m}(\lambda)
\nabla_{D_1}\cdots\nabla_{D_{m-1}} \mathcal{R}_{GD_m}(\lambda')\,.
\eea

These maps can be taken as constant at the source or image centre. Note that their dimensions are different because the image-to-source map is contracted with angular sizes, and while the source-to-image map is contracted with physical distances. Thus, the dimensions of the components of the maps are $[\mathcal{M}^{{}^\nwarrow}_{AB_1\cdots B_m}]\sim[\text{length}]^{-m}$ and $[\mathcal{M}^{{}^\searrow}_{AB_1\cdots B_m}]\sim[\text{length}]$.
\\[3mm]
\noindent{\bf Notation:} In what follows there is essentially no difference between the maps $(\nwarrow)$ and $(\searrow)$, so we use the generic notation
\be
\xi^A_{(m)} = \mathcal{M}^{A}_{~~B_1\cdots B_m}\zeta^{B_1}\cdots\zeta^{B_m}
\ee
with the understanding that if we are dealing with the image-to-source case $(\searrow)$ then $\bm\zeta$ represents the observed angle at the observer, and $\bm\xi_{(m)}$ the $m$'th order correction to the distance between two rays at the source (with an $m!$ factor). In the source-to-image case $(\nwarrow)$ we have $\bm\zeta$ representing the physical distance at the source (denoted $\bm\eta$ above), and $\bm\xi_{(m)}$ is the $m$'th order correction to the angle at the image.

\section{Invariant decomposition of a projected map}

Here we shall give a description of the decomposition of a generic map with no reference to lensing. We shall perform our decomposition on a plane although the extension to a sphere is relatively straightforward. (The lensing approximation we have used above already assumes a kind of flat sky approximation, because we have neglected everything except transverse derivatives in the screen space.) We perform a decomposition of the map in the plane in a real Cartesian basis, a polar basis, then consider the same in a helicity basis for completeness, which is where the relation between trace-free tensors and spin is most apparent. The calculation appears a little simpler in the latter, but the conversion back to real space removes this slight advantage.

\subsection{Decomposition of the map using symmetric trace-free tensors}

At each order the map $\mathcal{M}_{AB_1\cdots B_m}$ can be invariantly decomposed into a family of symmetric trace-free tensors which give the normal modes of each lensing mode. Note that we are only interested in $\mathcal{M}_{A(B_1\cdots B_m)}$ since the map is multiplied by the symmetric tensor $\zeta^{B_1}\cdots\zeta^{B_m}$ or $\eta^{B_1}\cdots\eta^{B_m}$ depending on the map in question (we will just use $\zeta^A$ from hereon as there is no real distinction between the 2 cases). The first split is to separate out the antisymmetric component from the first index:
\be
\mathcal{M}_{AB_1\cdots B_m} = \mathcal{M}_{AB_1\cdots B_m}^\text{symm}+\varepsilon_{A(B_1}{\mathcal{M}}_{B_2\cdots B_m)}^\text{anti-symm}
\ee
where $\mathcal{M}_{AB_1\cdots B_m}^\text{symm}=\mathcal{M}_{(AB_1\cdots B_m)}$ is symmetric as is ${\mathcal{M}}_{B_2\cdots B_m}^\text{anti-symm}={\mathcal{M}}_{(B_2\cdots B_m)}^\text{anti-symm}$, and $\epsilon_{AB}=-\epsilon_{BA}$ is the Levi-Civita tensor on the screen space.  Now, any symmetric tensor can be expanded into a sum of invariant STF tensors, which are the invariant normal modes of the map. For example, for linear weak lensing the split is into a scalar (convergence) and a rank-2 STF tensor (the shear).  The general expansion is a mess, but  the general idea is simple. For rank 4, we have
\be
\mathcal{M}_{ABCD}=\widehat\mathcal{M}\delta_{(AB}\delta_{CD)}+\widehat\mathcal{M}_{(AB}\delta_{CD)}+\widehat\mathcal{M}_{ABCD},
\ee
where a hat denotes that the tensor is trace-free (although the hat notation is redundant for rank-0 and -1 tensors, hats are placed for clarity). Similarly for a odd-ranked tensor, say rank 7, we have
\be
\mathcal{M}_{ABCDEFG}=\widehat\mathcal{M}_{(A}\delta_{BC}\delta_{DE}\delta_{FG)}+\widehat\mathcal{M}_{(ABC}\delta_{DE}\delta_{FG)}+\widehat\mathcal{M}_{(ABCDE}\delta_{FG)}+\widehat\mathcal{M}_{ABCDEFG}\,,
\ee
and so on. Note that each STF tensor has 2 degrees of freedom, and the number of indices represents the spin of the mode. First let us consider the case where $m$ is odd, so that the map itself is even:
\bea
\mathcal{M}_{AB_1\cdots B_m}^\text{symm}\hat\zeta^{B_1}\cdots\hat\zeta^{B_m}&=&
\bigg[\widehat\mathcal{M}\delta_{(AB_1}\cdots\delta_{B_{m-1}B_m)}
+\widehat\mathcal{M}_{(AB_1}\delta_{B_2B_3}\cdots\delta_{B_{m-1}B_m)}
+\cdots\nonumber\\
\cdots+
\widehat\mathcal{M}_{(AB_1\cdots B_{j}}\delta_{B_{j+1}B_{j+2}}\cdots\delta_{B_{m-1}B_m)}
&+&\cdots+
\widehat\mathcal{M}_{(AB_1\cdots B_{m-3}}\delta_{B_{m-1}B_m)}
+\widehat\mathcal{M}_{AB_1\cdots B_{m-1}B_m}
\bigg]\hat\zeta^{B_1}\cdots\hat\zeta^{B_m}\,,
\eea
where $j$ is odd. In the case where the map is odd, $m$ is even and we have instead
\bea
\mathcal{M}_{AB_1\cdots B_m}^\text{symm}\hat\zeta^{B_1}\cdots\hat\zeta^{B_m}&=&
\bigg[\widehat\mathcal{M}_{(A}\delta_{B_1B_2}\cdots\delta_{B_{m-1}B_m)}
+\widehat\mathcal{M}_{(AB_1B_2}\delta_{B_3B_4}\cdots\delta_{B_{m-1}B_m)}
+\cdots\nonumber\\
\cdots+
\widehat\mathcal{M}_{(AB_1\cdots B_{j}}\delta_{B_{j+1}B_{j+2}}\cdots\delta_{B_{m-1}B_m)}
&+&\cdots+
\widehat\mathcal{M}_{(AB_1\cdots B_{m-3}}\delta_{B_{m-1}B_m)}
+\widehat\mathcal{M}_{AB_1\cdots B_{m-1}B_m}
\bigg]\hat\zeta^{B_1}\cdots\hat\zeta^{B_m}\,,
\eea
where now $j\leq m$ is also even. 

Now consider the anti-symmetric component of the map. In the case where $m$ is odd we have 
\bea
\varepsilon_{A(B_1}{\mathcal{M}}_{B_2\cdots B_m)}^\text{anti-symm}\hat\zeta^{B_1}\cdots\hat\zeta^{B_m} &=&\varepsilon_{A(B_1}
\bigg[\overline{\mathcal{M}}\delta_{B_2B_3}\cdots\delta_{B_{m-1}B_m)}
+\overline{\mathcal{M}}_{B_2B_3}\delta_{B_4B_5}\cdots\delta_{B_{m-1}B_m)}
+\cdots\nonumber\\
\cdots+
\overline{\mathcal{M}}_{B_2B_3\cdots B_{j}}\delta_{B_{j+1}B_{j+2}}\cdots\delta_{B_{m-1}B_m)}
&+&\cdots+
\overline{\mathcal{M}}_{B_2B_3\cdots B_{m-3}}\delta_{B_{m-1}B_m)}
+\overline{\mathcal{M}}_{B_2B_3\cdots B_{m-1}B_m}
\bigg]\hat\zeta^{B_1}\cdots\hat\zeta^{B_m}\,,
\eea
and for $m$ even,
\bea
\varepsilon_{A(B_1}{\mathcal{M}}_{B_2\cdots B_m)}^\text{anti-symm}\hat\zeta^{B_1}\cdots\hat\zeta^{B_m} &=&\varepsilon_{A(B_1}
\bigg[\overline{\mathcal{M}}_{B_2}\delta_{B_3B_4}\cdots\delta_{B_{m-1}B_m)}
+\overline{\mathcal{M}}_{B_2B_3B_4}\delta_{B_5B_6}\cdots\delta_{B_{m-1}B_m)}
+\cdots\nonumber\\
\cdots+
\overline{\mathcal{M}}_{B_2B_3\cdots B_{j}}\delta_{B_{j+1}B_{j+2}}\cdots\delta_{B_{m-1}B_m)}
&+&\cdots+
\overline{\mathcal{M}}_{B_2B_3\cdots B_{m-3}}\delta_{B_{m-1}B_m)}
+\overline{\mathcal{M}}_{B_2B_3\cdots B_{m-1}B_m}
\bigg]\hat\zeta^{B_1}\cdots\hat\zeta^{B_m}\,.
\eea
Here we have used bars to represent trace free tensors which have come from ${\mathcal{M}}_{B_2\cdots B_m}^\text{anti-symm}$.

\subsection{The normal modes in the Cartesian basis}

 To see how these STF tensors relate to normal normal modes like shear and convergence, consider a point in the screen space given in Cartesian coordinates, $\zeta_A=r\hat\zeta_A$, where $\hat\zeta_A=(\cos\theta,\sin\theta)$ is the radial unit vector.  Two types of terms appear, those with all indices contracted with $\zeta$'s, and those with all but one contracted. Consider
 \be
 \widehat\mathcal{M}_{A_1\cdots A_n}\hat\zeta^{A_1}\cdots\hat\zeta^{A_n}
 = \sum_{k=0}^n {n\choose k} \cos^{n-k}\theta\sin^k\theta \widehat\mathcal{M}_{x^{n-k}y^k}\,,
 \ee
where a power on a tensor index means its repeated that many times. (This relation is easy to prove by induction where the binomial coefficients pop out naturally.)
Now let us define the two normal modes as determined by whether there are an odd or even number of $y$'s in $\mathcal{M}_{x^{n-k}y^k}$:
\bea
\alpha_n &=& (-1)^j \widehat{\mathcal{M}}_{x^{n-2j}y^{2j}} \\
\beta_n  &=& (-1)^j \widehat{\mathcal{M}}_{x^{n-2j-1}y^{2j+1}}\,, 
\eea
where the subscript on the normal modes carries the spin of the mode (we define $\beta_0=0$).
For each $n$ we can define a set of irreducible STF basis tensors 
\be
e_{A_1\cdots A_n}:~e_{x\cdots x}=1,~e_{x\cdots xy}=0~~~\mbox{and}~~~\tilde e_{A_1\cdots A_n}:~\tilde e_{x\cdots x}=0,~\tilde e_{x\cdots xy}=1\,,
\ee
and then
\be
\widehat\mathcal{M}_{A_1\cdots A_n} = \alpha_n\, e_{A_1\cdots A_n}+\beta_n \,\tilde e_{A_1\cdots A_n}\,.
\ee
(For example, for $n=2$ we have $e_{AB}=e^x_Ae^x_B-e^y_Ae^y_B$ and $\tilde e_{AB}=2e^x_{(A}e^y_{B)}$.) Note that $e_{x^{n-2j}y^{2j}}=(-1)^j$ and $\tilde e_{x^{n-2j-1}y^{2j+1}}=(-1)^j$. )
The total projection of the two STF basis tensors along $\hat{\bm\zeta}$ (i.e., when all indices are projected onto $\hat{\bm\zeta}$) are
\bea
e_{A_1\cdots A_n}\hat\zeta^{A_1}\cdots\hat\zeta^{A_n}&=&\sum_{k~\text{even}}^n {n\choose k} (-1)^{k/2}\cos^{n-k}\theta\sin^k\theta=\cos n\theta\,,\\
\tilde e_{A_1\cdots A_n}\hat\zeta^{A_1}\cdots\hat\zeta^{A_n}&=&\sum_{k~\text{odd}}^n {n\choose k} (-1)^{(k-1)/2}\cos^{n-k}\theta\sin^k\theta=\sin n\theta\,,
\eea
while for the case that one index remains unprojected, we have (using column vector notation for the unprojected index)
\bea
e_{A_1\cdots A_n}\hat\zeta^{A_2}\cdots\hat\zeta^{A_n}&=&
\sum_{k=0}^{n-1} {n-1\choose k} \cos^{n-1-k}\theta\sin^k\theta  {(-1)^{k/2}~~\text{if $k$ even, $0$ if $k$ odd}~~\choose(-1)^{(k+1)/2}~~\text{if $k$ odd, $0$ if $k$ even}} 
= {\cos(n-1)\theta\choose -\sin(n-1)\theta}\,,\\
\tilde e_{A_1\cdots A_n}\hat\zeta^{A_2}\cdots\hat\zeta^{A_n}&=&
\sum_{k=0}^{n-1} {n-1\choose k} \cos^{n-1-k}\theta\sin^k\theta  {(-1)^{(k-1)/2}~~\text{if $k$ odd, $0$ if $k$ even}\choose(-1)^{k/2}~~\text{if $k$ even, $0$ if $k$ odd}} 
= {\sin(n-1)\theta\choose \cos(n-1)\theta}\,.
\eea
From these key relations we can find all the possible projections required for the lensing map.

We have
\bea
 \widehat\mathcal{M}_{A_1\cdots A_n}\hat\zeta^{A_1}\cdots\hat\zeta^{A_n}
&=& \alpha_n\,\cos(n\theta)+\beta_n\,\sin(n\theta)\,.
\eea
and
\bea
 \widehat\mathcal{M}_{A_1A_2\cdots A_n}\hat\zeta^{A_2}\cdots\hat\zeta^{A_n}
&=&\left(\begin{array}{cc}\alpha_n & \beta_n \\\beta_n & -\alpha_n\end{array}\right)
{\cos(n-1)\theta\choose \sin(n-1)\theta}\,.
\eea

We can now perform a full split of the symmetric part of $\widehat\mathcal{M}_{AB_1\cdots B_m}$, as it appears in the general lensing map. 
The generic term becomes
\bea
\widehat\mathcal{M}_{(AB_1\cdots B_{j}}\delta_{B_{j+1}B_{j+2}}\!\!\!&\cdots&\!\!\!\delta_{B_{m-1}B_m)}\hat\zeta^{B_1}\cdots\hat\zeta^{B_m}=
\frac{j+1}{m+1}\widehat\mathcal{M}_{AB_1\cdots B_{j}}\hat\zeta^{B_1}\cdots\hat\zeta^{B_j}+\frac{m-j}{m+1}\hat\zeta_A\,\widehat\mathcal{M}_{B_1\cdots B_{j+1}}\hat\zeta^{B_1}\cdots\hat\zeta^{B_{j+1}}
\nonumber\\
&=&\frac{m+j+2}{2(m+1)}\bigg[
\alpha_{j+1}^m\left(\begin{array}{cc}1 & 0 \\0 & -1\end{array}\right)
+\beta_{j+1}^m\left(\begin{array}{cc}0 & 1 \\1 & 0\end{array}\right)
\bigg]{\cos j\theta\choose\sin j\theta}\nonumber\\&&
\frac{m-j}{2(m+1)}\bigg[
\alpha_{j+1}^m\left(\begin{array}{cc}1 & 0 \\0 & 1\end{array}\right)
+\beta_{j+1}^m\left(\begin{array}{cc}0 & 1 \\-1 & 0\end{array}\right)
\bigg]{\cos (j+2)\theta\choose\sin (j+2)\theta}
\eea
We have added an $m$ super-script to the normal mode amplitudes to denote the fact that they arise in a tensor with $m$ indices contracted as on the lhs (this will be useful below).
This formula is for all $j$, and actually works with $j=-1$ as well if we take it to mean the $\widehat\mathcal{M}$ term in the even case, with $\widehat\mathcal{M}=\alpha^m_{0}$. 

With this we can now write, for any $m$,
\bea
\mathcal{M}_{AB_1\cdots B_m}^\text{symm}\hat\zeta^{B_1}\cdots\hat\zeta^{B_m}&=&
\sum_{s=0}^{m+1} \frac{[1-(-1)^{m+s}]}{4(m+1)}\bigg\{
(m+s+1)\bigg[
\alpha_{s}^m\left(\begin{array}{cc}1 & 0 \\0 & -1\end{array}\right)
+\beta_{s}^m\left(\begin{array}{cc}0 & 1 \\1 & 0\end{array}\right)
\bigg]{\cos (s-1)\theta\choose\sin (s-1)\theta}\nonumber\\&&
~~~~~~~~~~~~~~~~~~~~~~~~~~(m-s+1)\bigg[
\alpha_{s}^m\left(\begin{array}{cc}1 & 0 \\0 & 1\end{array}\right)
+\beta_{s}^m\left(\begin{array}{cc}0 & 1 \\-1 & 0\end{array}\right)
\bigg]{\cos (s+1)\theta\choose\sin (s+1)\theta}\bigg\}
\eea
(Note that the factor $[1-(-1)^{m+s}]/2$ selects the terms such that $m+s$ is odd.) We have labeled the terms in the sum such that $s$ carries the spin of the mode. 

The general term in the antisymmetric component of the map becomes, using $\varepsilon_{AB}\hat\zeta^{B}=(\sin\theta,-\cos\theta)$,
\bea
\varepsilon_{AB_1}\overline{\mathcal{M}}_{B_2B_3\cdots B_{j}}\delta_{B_{j+1}B_{j+2}}\cdots\delta_{B_{m-1}B_m}\hat\zeta^{B_1}\cdots\hat\zeta^{B_m}&=& \varepsilon_{AB_1}\hat\zeta^{B_1}\overline{\mathcal{M}}_{B_2B_3\cdots B_{j}}\hat\zeta^{B_2}\cdots\hat\zeta^{B_j}\nonumber\\
&=&\frac{1}{2}\bigg[
-\bar\alpha_{j-1}^m\left(\begin{array}{cc}0 & 1 \\1 & 0\end{array}\right)
+\bar\beta_{j-1}^m\left(\begin{array}{cc}1 & 0 \\0 &  -1\end{array}\right)
\bigg]{\cos (j-2)\theta\choose\sin (j-2)\theta}\nonumber\\&&
+\frac{1}{2}\bigg[
\bar\alpha_{j-1}^m\left(\begin{array}{cc}0 & 1 \\-1 & 0\end{array}\right)
-\bar\beta_{j-1}^m\left(\begin{array}{cc}1 & 0 \\0 & 1\end{array}\right)
\bigg]{\cos j\theta\choose\sin j\theta}\,.
\eea
This formula works for $j=1$ with $\overline{\mathcal{M}}=\alpha^m_1$.
Consequently we have, similarly to the symmetric case above,
\bea
\varepsilon_{A(B_1}{\mathcal{M}}_{B_2\cdots B_m)}^\text{anti-symm}\hat\zeta^{B_1}\cdots\hat\zeta^{B_m} &=&
\sum_{s=0}^{m-1} \frac{[1-(-1)^{m+s}]}{4}\bigg\{
\bigg[
-\bar\alpha_{s}^m\left(\begin{array}{cc}0 & 1 \\1 & 0\end{array}\right)
+\bar\beta_{s}^m\left(\begin{array}{cc}1 & 0 \\0 &  -1\end{array}\right)
\bigg]{\cos (s-1)\theta\choose\sin (s-1)\theta}\nonumber\\&&
+\bigg[
\bar\alpha_{s}^m\left(\begin{array}{cc}0 & 1 \\-1 & 0\end{array}\right)
-\bar\beta_{s}^m\left(\begin{array}{cc}1 & 0 \\0 & 1\end{array}\right)
\bigg]{\cos (s+1)\theta\choose\sin (s+1)\theta}
\bigg\}\,.
\eea
Note that $\bar\alpha_{s}^m$ and $\bar\beta_{s}^m$ are not truly independent modes as they always appear in the combination $\bar\alpha_{s}^m\cos s\theta+\bar\beta_{s}^m\sin s\theta$ (this is the `radial mode' of the tensor).

Our final map at $m$'th order becomes
\bea
&&\mathcal{M}_{AB_1\cdots B_m}\hat\zeta^{B_1}\cdots\hat\zeta^{B_m}=
\sum_{s=0}^{m+1} \frac{[1-(-1)^{m+s}]}{4}\bigg\{
\nonumber\\&&
\bigg[
\bigg[\bigg(1+\frac{s}{m+1}\bigg)
\alpha_{s}^m+\bar\beta_s^m\bigg]\left(\begin{array}{cc}1 & 0 \\0 & -1\end{array}\right)
+\bigg[-\bar\alpha_{s}^m
+\bigg(1+\frac{s}{m+1}\bigg)\beta_{s}^m\bigg]\left(\begin{array}{cc}0 & 1 \\1 & 0\end{array}\right)
\bigg]{\cos (s-1)\theta\choose\sin (s-1)\theta}
\nonumber\\&&
\bigg[
\bigg[\bigg(1-\frac{s}{m+1}\bigg)
\alpha_{s}^m-\bar\beta^m_s\bigg]\left(\begin{array}{cc}1 & 0 \\0 & 1\end{array}\right)~~
+\bigg[+\bar\alpha_{s}^m
+\bigg(1-\frac{s}{m+1}\bigg)\beta_{s}^m\bigg]\left(\begin{array}{cc}0 & 1 \\-1 & 0\end{array}\right)
\bigg]{\cos (s+1)\theta\choose\sin (s+1)\theta}\bigg\}\,,
\label{jhdfbvdfhbvdfbb}
\eea
where we define
\be
\bar\alpha_{m}^m=\bar\alpha_{m+1}^m=\bar\beta_{m}^m=\bar\beta_{m+1}^m=0\,.
\ee
Other $\alpha^m_s$'s and $\beta^m_s$'s which do not appear in the sum (i.e., if $m+s$ is even) can be assumed to be zero. 

\subsection{The normal modes in the  polar basis}

A polar basis at angle $\theta$ is
\bea
\bm e_r&=&\cos\theta\bm e_x+\sin\theta\bm e_y\,,\nonumber\\
\bm e_\theta&=&-\sin\theta\bm e_x+\cos\theta\bm e_y
\eea
so that $\hat{\bm \zeta}=\bm e_r$. We can define a set of radial and angular STF basis tensors as
\be
e^{(r)}_{A_1\cdots A_n}:~e^{(r)}_{r\cdots r}=1,~e^{(r)}_{r\cdots r\theta}=0~~~\mbox{and}~~~e^{(\theta)}_{A_1\cdots A_n}:~ e^{(\theta)}_{r\cdots r}=0,~ e^{(\theta)}_{r\cdots r\theta}=1\,,
\ee
and then for a general STF tensor,
\be
\widehat\mathcal{M}_{A_1\cdots A_n} = \sigma_n\, e^{(r)}_{A_1\cdots A_n}+\omega_n \, e^{(\theta)}_{A_1\cdots A_n}\,,
\ee
where
\bea
\sigma_n&=&\alpha_n\,\cos n\theta+\beta_n\,\sin n\theta\,,\nonumber\\
\omega_n&=&-\alpha_n\,\sin n\theta+\beta_n\,\cos n\theta\,.
\eea
That is, the independent degrees of freedom in the polar basis are those in the Cartesian basis, but `rotated' by $n\theta$. 
For the projections of an STF tensor with $\bm\zeta$ appearing in the general map we have
\bea
 \widehat\mathcal{M}_{A_1\cdots A_s}\hat\zeta^{A_1}\cdots\hat\zeta^{A_s}
 &=&\sigma_s
\eea
and
\bea
 \widehat\mathcal{M}_{A_1\cdots A_s}\hat\zeta^{A_2}\cdots\hat\zeta^{A_s}
 &=& \sigma_s\,\bm e_r+\omega_s\,\bm e_\theta\,.
\eea
The generic terms become
\bea
\widehat\mathcal{M}_{(AB_1\cdots B_{s-1}}\delta_{B_{s}B_{s+1}}\cdots\delta_{B_{m-1}B_m)}\hat\zeta^{B_1}\cdots\hat\zeta^{B_m}
&=& \sigma_{s}\bm e_r+\frac{s}{m+1}\omega_{s}\bm e_\theta\\
\varepsilon_{AB_1}\overline{\mathcal{M}}_{B_2B_3\cdots B_{s+1}}\delta_{B_{s+2}B_{s+3}}\cdots\delta_{B_{m-1}B_m}\hat\zeta^{B_1}\cdots\hat\zeta^{B_m}&=&-\overline{\sigma}_{s}\bm e_\theta
\eea
The total map is then
\bea
&&\mathcal{M}_{AB_1\cdots B_m}\hat\zeta^{B_1}\cdots\hat\zeta^{B_m}=
\sum_{s=0}^{m+1} \frac{[1-(-1)^{m+s}]}{2}\bigg\{\sigma_s^m\bm e_r+\left(\frac{s}{m+1}\omega_{s}^m-\overline{\sigma}_{s}^m\right)\bm e_\theta
\bigg\}
\eea
with $\overline{\sigma}_{m}^m=\overline{\sigma}_{m+1}^m=0$

\subsection{The normal modes in the helicity basis}\label{djkscskjcnsk}

An alternative basis for the analysis is the helicity basis defined by the complex null vectors~\cite{Castro:2005bg}
\be
    \bm e_\pm=\frac{1}{\sqrt{2}}\left(\bm e_x\pm \i\bm e_y\right) =\frac{e^{\pm\i\theta}}{\sqrt{2}}\left(\bm e_r\pm \i\bm e_\theta\right)  \,,
\ee
where 
\be
\bm e_\pm\cdot \bm e_\pm=0,~~~~\bm e_\pm\cdot \bm e_\mp=1~~~\mbox{and}~~~\bm e_x=\frac{1}{\sqrt{2}}\left(\bm e_++\bm e_-\right)\,,~~~\bm e_y=\frac{\i}{\sqrt{2}}\left(\bm e_--\bm e_+\right)\,.
\ee
The metric and Levi-Civita tensor are
\bea
g_{AB}&=& 2e^+_{(A}e^-_{B)} = e_A^xe_B^x+e_A^ye_B^y\,,\\
\varepsilon_{AB}&=&2\i \,e^+_{[A}e^-_{B]} = 2e^x_{[A}e^y_{B]}\,.
\eea
In this basis the unit radial vector on the screen space becomes
\be\label{dksjnskjvnskd}
\hat{\bm\zeta}=\cos\theta\bm e_x+\sin\theta\bm e_y = \frac{1}{\sqrt{2}}\left(e^{-\i\theta}\bm e_++e^{\i\theta}\bm e_- \right) = {}_-\hat\zeta\bm e_++{}_+\hat\zeta\bm e_-\,.
\ee
In the helicity basis we  write the components of $\hat{\bm\zeta}$ as 
\be
\hat\zeta^\mp={}_\pm\hat\zeta=\hat{\bm\zeta}\cdot\bm e_\pm=e^{\pm\i\theta}/\sqrt{2}\,.
\ee
Note that if we lower the index we have $\hat\zeta_\pm=\hat\zeta^\mp={}_\pm\hat\zeta$.

Under a counterclockwise rotation by $\psi$ to a new basis $\bm e_A'=R(\psi)_{A}^{~B}\bm e_B$, where $R(\psi)_{A}^{~B}$ is the rotation matrix, the helicity basis changes by 
\be\label{odfvpfjvdpfokvpd}
\bm e_\pm'=e^{\pm\i\psi}\bm e_\pm\,.
\ee
 Because the vector $\hat{\bm\zeta}$ is invariant under this rotation, the components in the helicity basis must change contrary to the basis, giving
\be
{}_\pm\hat\zeta'=\hat{\bm\zeta}\cdot\bm e_\pm'=e^{\pm\i\psi}\hat{\bm\zeta}\cdot\bm e_\pm=e^{\pm\i\psi}{}_\pm\hat\zeta\,.
\ee
The components of $\hat{\bm\zeta}$ in the helicity basis are said to have spin weight $\pm1$ because of this transformation property (which accounts for the slightly perverse notation in \ref{dksjnskjvnskd}).

In this basis symmetric, trace-free tensors have a particularly simple representation. Since $e_+^Ae_-^B\widehat{\mathcal{M}}_{\cdots A \cdots B \cdots} =~$trace terms$~=0$, the two independent components are
\be
{}_{\pm s}\widehat{\mathcal{M}}=  \widehat{\mathcal{M}}_{A_1\cdots A_s} e_\pm^{A_1}\cdots e_\pm^{A_s}\,,
\ee
and transform as spin $\pm s$ quantities from \eqref{odfvpfjvdpfokvpd} (and the fact that the tensor $\widehat{\mathcal{M}}_{A_1\cdots A_s}$ is invariant). The real space tensor can then be written in terms of the complex helicity basis as
\be
\widehat{\mathcal{M}}_{A_1\cdots A_s} = 
{}_{-s}\widehat{\mathcal{M}}\,\,e^+_{A_1}\cdots e^+_{A_s}
+{}_{+s}\widehat{\mathcal{M}}\,\,e^-_{A_1}\cdots e^-_{A_s}\,.
\ee
From this we find that the two independent modes in the Cartesian basis, $\alpha_s$ and $\beta_s$ are related to the spin-modes ${}_{\pm s}\widehat{\mathcal{M}}$ in the helicity basis via
\be
{}_{\pm s}\widehat{\mathcal{M}} = 2^{s/2-1}\left(\alpha_s\pm\i\beta_s\right)= 2^{s/2-1}e^{\pm\i s\theta}\left(\sigma_s\pm\i\omega_s\right)\,.
\ee

For the projections of an STF tensor with $\bm\zeta$ appearing in the general map we have
\bea
 \widehat\mathcal{M}_{A_1\cdots A_s}\hat\zeta^{A_1}\cdots\hat\zeta^{A_s}
 &=&2^{-s/2}\left({}_{-s}\widehat{\mathcal{M}}e^{+\i s\theta}+ {}_{+s}\widehat{\mathcal{M}}e^{-\i s\theta}\right)
\eea
and
\bea
 \widehat\mathcal{M}_{A_1\cdots A_s}\hat\zeta^{A_2}\cdots\hat\zeta^{A_s}
 &=&2^{-(s-1)/2}\left(
 {}_{-s}\widehat{\mathcal{M}}e^{+\i (s-1)\theta} \,e_{A_1}^++ {}_{+s}\widehat{\mathcal{M}}e^{-\i (s-1)\theta}\, e_{A_1}^-\right)\\
 &=& \frac{1}{2^{(s-1)/2}}
 \left(\begin{array}{cc}
 {}_{-s}\widehat{\mathcal{M}} & {}_{+s}\widehat{\mathcal{M}} 
 \\
 {}_{-s}\widehat{\mathcal{M}}\i & -{}_{+s}\widehat{\mathcal{M}}\i\end{array}\right)
 {e^{+\i (s-1)\theta}\choose e^{-\i (s-1)\theta}}\,.
\eea
In the last line the resulting column vector corresponds to the real space Cartesian basis, not the helicity basis. 
For the generic term in the symmetric part of the lensing map we have
\bea
\widehat\mathcal{M}_{(AB_1\cdots B_{s-1}}\delta_{B_{s}B_{s+1}}\!\!\!&\cdots&\!\!\!\delta_{B_{m-1}B_m)}\hat\zeta^{B_1}\cdots\hat\zeta^{B_m}=
\frac{s}{m+1}\widehat\mathcal{M}_{AB_1\cdots B_{s-1}}\hat\zeta^{B_1}\cdots\hat\zeta^{B_{s-1}}+\frac{m+1-s}{m+1}\hat\zeta_A\,\widehat\mathcal{M}_{B_1\cdots B_{s}}\hat\zeta^{B_1}\cdots\hat\zeta^{B_{s}}
\nonumber\\
&=&\frac{1}{2^{(s+1)/2}}\bigg\{
\left[ \left(1+\frac{s}{m+1}\right){}_{-s}\widehat{\mathcal{M}}e^{+\i (s-1)\theta}
+\left(1-\frac{s}{m+1}\right){}_{+s}\widehat{\mathcal{M}}e^{-\i (s+1)\theta}\right]e^+_A\nonumber\\
&& +\left[ \left(1-\frac{s}{m+1}\right){}_{-s}\widehat{\mathcal{M}}e^{+\i (s+1)\theta}
+\left(1+\frac{s}{m+1}\right){}_{+s}\widehat{\mathcal{M}}e^{-\i (s-1)\theta}\right]e^-_A\,.
\eea
The general term in the antisymmetric component of the map becomes, using $\varepsilon_{AB}\hat\zeta^{B}=\i/\sqrt{2}(e^{-\i\theta}e^+_A-e^{\i\theta}e^-_A)$,
\bea
\varepsilon_{AB_1}\overline{\mathcal{M}}_{B_2B_3\cdots B_{s+1}}\delta_{B_{s+2}B_{s+3}}\cdots\delta_{B_{m-1}B_m}\hat\zeta^{B_1}\cdots\hat\zeta^{B_m}&=& \varepsilon_{AB_1}\hat\zeta^{B_1}\overline{\mathcal{M}}_{B_2B_3\cdots B_{s+1}}\hat\zeta^{B_2}\cdots\hat\zeta^{B_{s+1}}\nonumber\\
=\frac{\i}{2^{(s+1)/2}}\bigg\{
\left[ {}_{-s}\overline{\mathcal{M}}e^{+\i (s-1)\theta}
+{}_{+s}\overline{\mathcal{M}}e^{-\i (s+1)\theta}\right]e^+_A
& -&\left[{}_{-s}\overline{\mathcal{M}}e^{+\i (s+1)\theta}
+{}_{+s}\overline{\mathcal{M}}e^{-\i (s-1)\theta}\right]e^-_A\bigg\}\,.
\eea
A sum over the spin modes of the last two projections gives us~\eqref{jhdfbvdfhbvdfbb} in the helicity basis.

\subsection{Relation to convergence, shear and flexion}

Let us check this against known results. For $m=1$ we have simply the Jaccobi map, which, in this notation has contributions from $s=0$ and $s=2$ modes:
\bea
\mathcal{M}_{AB}\zeta^{B}&=&r 
\bigg[
\alpha_{0}^1\left(\begin{array}{cc}1 & 0 \\0 & 1\end{array}\right)
+\bar\alpha_{0}^1
\left(\begin{array}{cc}0 & 1 \\-1 & 0\end{array}\right)
+
\alpha_{2}^1\left(\begin{array}{cc}1 & 0 \\0 & -1\end{array}\right)
+\beta_{2}^1\left(\begin{array}{cc}0 & 1 \\1 & 0\end{array}\right)
\bigg]{\cos \theta\choose\sin \theta}\,.
\eea
Comparing with the usual amplification matrix we see $\alpha_0^1$ is (minus) the convergence, $\bar\alpha_0^1$ is the rotation, and $\alpha_{2}^1$ and $\beta_{2}^1$ are the 2 independent polarizations of the shear. In the radial basis we have
\bea
\mathcal{M}_{AB}\zeta^{B}&=&r 
\big[(\sigma_0^1+\sigma_2^1)\bm e_r+(-\overline\sigma_0^1+\omega_2^1)\bm e_\theta
\big]\,.
\eea
The second-order map becomes
\bea
&&\mathcal{M}_{ABC}\zeta^{B}\zeta^{C}=\frac{1}{2}r^2\bigg\{
\left(\frac{4}{3}\alpha_1^2+\bar\beta_1^2\right){1\choose0}
+\left(-\bar\alpha_1^2+\frac{4}{3}\beta_1^2\right){0\choose1}
\nonumber\\&&
+\bigg[
\left(\frac{2}{3}\alpha_1^2-\bar\beta_1^2\right)\left(\begin{array}{cc}1 & 0 \\0 &  1\end{array}\right)
+\left(\bar\alpha_1^2+\frac{2}{3}\beta_1^2\right)\left(\begin{array}{cc}0 & 1 \\-1 &  0\end{array}\right)
+2\alpha_3^2\left(\begin{array}{cc}1 & 0 \\0 &  -1\end{array}\right)
+2\beta_3^2\left(\begin{array}{cc}0 & 1 \\1 &  0\end{array}\right)
\bigg]{\cos 2\theta\choose\sin2\theta}
\bigg\}\,.
\eea
The first two terms are just a shift in the centre of the source, while the next two correspond to spin-1 $\mathcal{F}$-type flexion, and the last two spin-3 $\mathcal{G}$-type flexion.

\subsection{Action of the normal modes~-- roulettes }

\begin{SCfigure}
~~~~\includegraphics[width=0.7\textwidth]{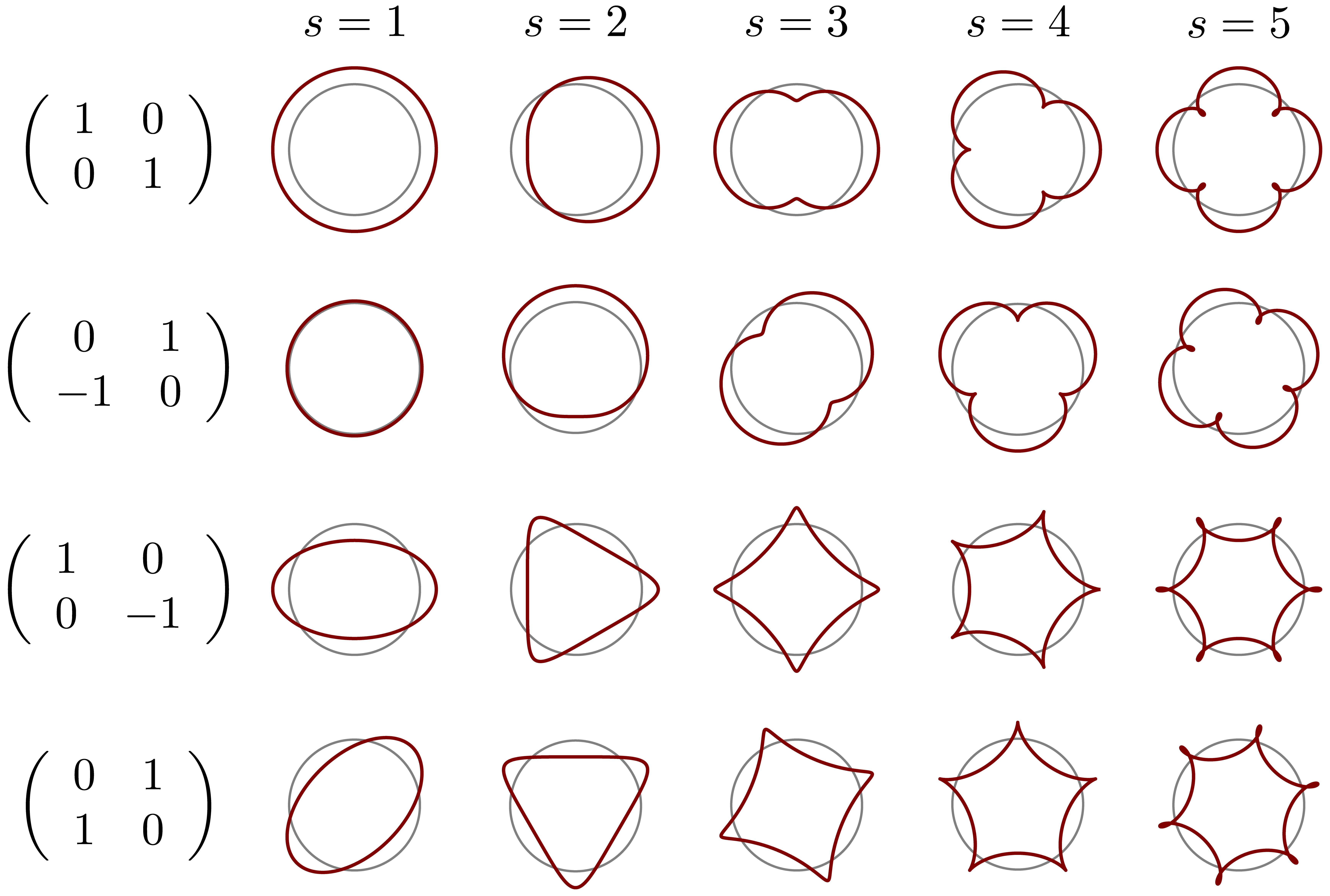}~~~~~~~~
\caption{Distortions of the unit circle induced by the rotation and reflection matrices $\bm I, \bm \varepsilon, \bm R_-, \bm R_/$, from top to bottom. Here we have plotted $\hat{\bm\zeta}+0.25\,\bm R\,\cdot\,\bm p_{(s)}$\,, where $\bm R$ is one of $\bm I, \bm \varepsilon, \bm R_-, \bm R_/$.  The matrices $\bm I, \bm \varepsilon$ trace out the same epitrochoid of rotational spin $s-1$, rotated by angles $\pi/2(s-1)$ from each other. The matrices $\bm R_-,\bm R_/$ trace out hypotrochoids of spin $s+1$, rotated by $\pi/2(s+1)$.}
\label{sdkjbhvdfhj}
\end{SCfigure}
Let us now investigate each type of  normal mode which appears. This is simplest in the Cartesian basis. Generically we consider the action of a mode as a distortion of the unit circle. Consider the distortion induced by a vector 
\be
\e_{(s)}=\cos s\theta\,\,\bm e_x+\sin s\theta\,\,\bm e_y=\cos (s-1)\theta\,\,\bm e_r+\sin (s-1)\theta\,\,\bm e_\theta\,.
\ee
This vector traces out a curve with spin $s-1$.
 The 4 independent distortions associated with this mode which appear in the Cartesian basis in~\eqref{jhdfbvdfhbvdfbb} are given by the orthogonal matrices
\be
\bm I          = \left(\begin{array}{cc}1 & 0 \\0 & 1\end{array}\right)\,,~~~
\bm\varepsilon= \left(\begin{array}{cc}0 & 1 \\-1 & 0\end{array}\right)\,,~~~
\bm R_-         = \left(\begin{array}{cc}1 & 0 \\0 & -1\end{array}\right)\,,~~~
\bm R_/          = \left(\begin{array}{cc}0 & 1 \\1 & 0\end{array}\right)\,,
\ee 
The first two are rotation matrices (with unit determinant) by $0$ and $-\pi/2$ respectively,  and the second two are reflections in the  line $y=x$ and $x$-axis  (with determinant $-1$). These are the Pauli spin matrices with $\varepsilon=\i\sigma_2$, and so obey similar commutation relations. 
We show plots of the distortions these induce to the unit circle in Fig.~\ref{sdkjbhvdfhj}, using
\be
\hat{\bm\zeta} + \alpha \bm R\cdot{\e}_{(s)}\,,
\ee
where $\alpha$ is a small constant amplitude. The length of this vector varies as
\be
|\hat{\bm\zeta} + \alpha \bm R\cdot{\e}_{(s)}|^2= 1+\alpha^2+2\alpha\left\{
\begin{array}{l|l}
\cos(s-1)\theta    & \bm R=\bm I  \\
\sin(s-1)\theta    &  \bm R=\bm\varepsilon  \\
\cos(s+1)\theta     &  \bm R=\bm R_-  \\
 \sin(s+1)\theta  &  \bm R=\bm R_/
\end{array}
\right.\,,
\ee
so the rotation matrices give rise to a spin $s-1$ shape, and the reflection matrices give rise to spin $s+1$ shapes. In fact, these are all trochoids\footnote{From \href{https://en.wikipedia.org/wiki/Centered_trochoid}{Wikipedia}: A centred trochoid is the roulette formed by a circle rolling along another circle. An epitrochoid is a roulette traced by a point attached to a circle of radius r rolling around the outside of a fixed circle of radius R, where the point is a distance d from the centre of the exterior circle. A hypotrochoid is a roulette traced by a point attached to a circle of radius r rolling around the inside of a fixed circle of radius R, where the point is a distance d from the centre of the interior circle.}~-- the first two are epitrochoids (formed from a roulette from a circle rolling outside a circle), and the other two are hypotrochoids (circle rolling inside a circle)~\cite{pawley}. In linear weak lensing, $\bm I$ and $\bm R$ give convergence and rotation, and $\bm R_-$ and $\bm R_/$ give the 2 components of the shear.

\begin{figure}[htbp]
\begin{center}
\includegraphics[width=\textwidth]{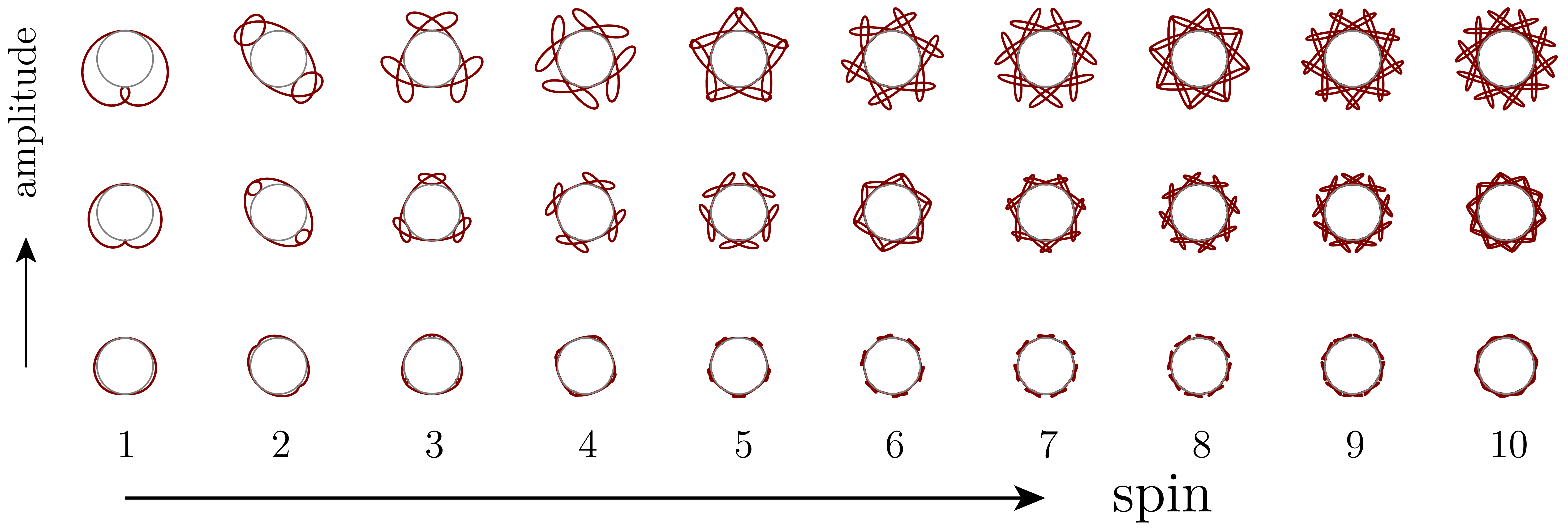}
\caption{The odd modes as a function of $\bar \alpha_s^m$. The $\bar \beta_s^m$ modes are the same, just rotated by $\pi/2s$}
\label{ldskncskjcnskj}
\end{center}
\end{figure}

Now lets move to the normal modes appearing in the map: the `even' modes $\alpha^m_s, \beta^m_s$ and the `odd' modes $\bar\alpha^m_s, \bar\beta^m_s$. We shall start with the odd modes as they are slightly simpler. Consider the $\bar\alpha_s^m$ mode, which distorts a unit circle in two parts from a reflection and a rotation:
\be
\hat{\bm\zeta}\mapsto \hat{\bm\zeta} + \bar\alpha_s^m\left(  -\bm R_/\cdot{\e}_{(s-1)} +\bm R\cdot {\e}_{(s+1)}
\right)
\ee
The first correction traces out a hypotrochoid, and the second correction an epitrochoid giving rise to an odd roulette. Both these curves have $s$ lobes, corresponding to a spin-$s$ mode. These modes are shown in Fig.~\ref{ldskncskjcnskj}. Note that the $\bar\beta^m_s$ mode is the same but rotated by $\pi/2s$. Referring to Fig.~\ref{sdkjbhvdfhj}, we can pick out the two modes giving rise to the odd distortion. 
For example, for $s=3$ we have a sum of the triangular shape in the bottom row, second column, and the 3-lobed shape two rows above. 

\begin{figure}[htbp]
\begin{center}
\includegraphics[width=\textwidth]{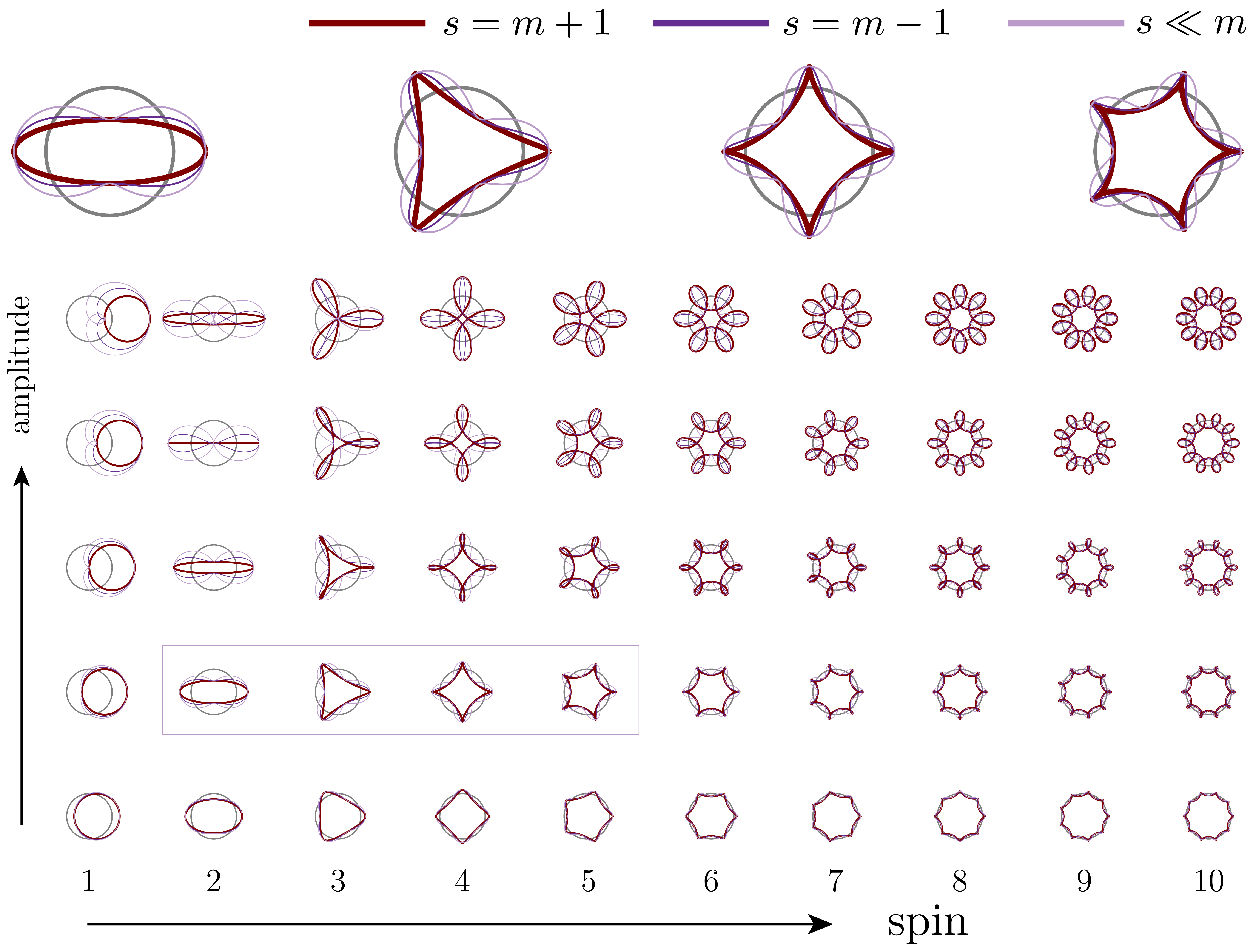}
\caption{The even modes as a function of $ \alpha_s^m$. The $ \beta_s^m$ modes are the same, just rotated by $\pi/2s$. For a fixed $s$, 3 $m$ values are considered, as the ratio between $s$ and $m+1$ gives quite different normal modes. The 4 enlarged figures at the top are a blow-up of the 4 in the rectangle, shown for clarity. }
\label{sjhdbcshvbfwoie}
\end{center}
\end{figure}

For the even modes, we again focus on the $\alpha^m_s$ modes, the $\beta^m_s$ being rotated by $\pi/2s$. We show these in Fig.~\ref{sjhdbcshvbfwoie}.
The even modes are more complicated because they appear in the map with a different weighting between the epi- and hypo-trochoid contributions. Here the roulettes are the distortion of the unit circle given by
\be
\hat{\bm\zeta}\mapsto \hat{\bm\zeta} + \alpha_s^m\left[  \bigg(1+\frac{s}{m+1}\bigg)\bm R_-\cdot{\e}_{(s-1)} +
\bigg(1-\frac{s}{m+1}\bigg)\bm I\cdot {\e}_{(s+1)}
\right]
\ee
Recall that the relative sizes of $s$ and $m$ occur from splitting off the traces of the total map. 
For a fixed $m$,  $m\gg s$ gives the mode which is mainly traces, whereas $s=m+1$ is the part of the map which is entirely trace-free. We show a selection in Fig.~\ref{sjhdbcshvbfwoie}~-- the two extremes, and the spin mode with one trace removed from the total map $s=m-1$.

\subsection{Finding the roulette amplitudes}

We can invert the expansion of the map into a sum over normal modes to give the coefficients $\alpha_s, \beta_s$ and $\bar\alpha_s, \bar\beta_s$ in terms of the projected map $\hat\xi^{(m)}_A=\mathcal{M}_{AB_1\cdots B_m}\hat\zeta^{B_1}\cdots\hat\zeta^{B_m}$, assuming that the coefficients are constant. This then becomes like a 2D Fourier series. Note that we can extract the $m$'th map from the full $\bm\xi$ just using 
\be
\hat\xi^{(m)}_A=\frac{1}{m!}\frac{\partial^m}{\partial r^m}\xi_A\bigg|_{r=0}\,.
\ee
 We write this in column vector form for the free index $A$ as $\hat{\bm\xi}_{(m)}={\bm{\mathcal{M}}}_{B_1\cdots B_m}\hat\zeta^{B_1}\cdots\hat\zeta^{B_m}$. (The $\hat{~}$ on the   $\xi$ denotes we have removed the factor of $r^m$ which appears in $\xi^{(m)}_A$.) Multiplying this on the right with the row vector $\e_{(n)}^T$ and integrating around the unit circle allows us to extract each spin mode, using
\be
\frac{1}{\pi}\int_{-\pi}^{\pi}\mathrm{d}\theta\,\,\e_{(m)}\e_{(n)}^T=
\left\{
\begin{array}{lcr}
\bm I\delta_{m,n} &\text{for}& m,n>0\\
\bm R_-\delta_{|m|,n} &\text{for}& m<0,n>0\\
\left(\begin{array}{cc}2 & 0 \\0 & 0\end{array}\right)\delta_{0,n}&\text{for}& m=0
\end{array}
\right.\,.
\ee
 In \eqref{jhdfbvdfhbvdfbb} this gives, for $m,n>0$,
\bea
&&\frac{1}{\pi}\int_{-\pi}^{\pi}\mathrm{d}\theta\,\,\hat{\bm\xi}_{(m)}\e_{(n)}^T=
%
\frac{1}{2}\left[A_{n-1}^-\bm I+B_{n-1}^-\bm \varepsilon+A_{n+1}^+\bm R_-+B_{n+1}^+\bm R_/
\right]\,,
\label{jhdfbvdfhscsdbvdfbb}
\eea
where
\bea
A_s^\pm=\bigg(1\pm\frac{s}{m+1}\bigg)\alpha_{s}^m\pm\bar\beta_s^m,~~~~B_s^\pm=\mp\bar\alpha_{s}^m+\bigg(1\pm\frac{s}{m+1}\bigg)\beta_{s}^m\,.
\eea
Using the fact that the matrices are orthogonal allows us to solve for the coefficients easily: we multiply~\eqref{jhdfbvdfhscsdbvdfbb} by each of the matrices in turn and take a trace to isolate each component~-- e.g., for $n+m$ even,  
\bea
A_{n+1}^+=\frac{1}{\pi}\int_{-\pi}^{\pi}\mathrm{d}\theta\,\,\mathrm{tr}\, (\hat{\bm\xi}_{(m)}\cdot\bm R_-\cdot\e_{(n)}^T)=\frac{1}{\pi}\int_{-\pi}^{\pi}\mathrm{d}\theta\,\, R^-_{AB}\hat{\xi}_{(m)}^A\pe_{(n)}^B\,.
\eea
We then have
\bea
\alpha_s^m&=&\frac{1}{2\pi}\int_{-\pi}^{\pi}\mathrm{d}\theta\,\,\hat{\xi}_{(m)}^A\left[ \pe^{(s+1)}_A+
 R^-_{AB}\pe_{(s-1)}^B \right]\,,\\
\beta_s^m&=& \frac{1}{2\pi}\int_{-\pi}^{\pi}\mathrm{d}\theta\,\,\hat{\xi}_{(m)}^A\left[ \varepsilon_{AB}\pe_{(s+1)}^B+
 R^/_{AB}\pe_{(s-1)}^B \right]\,,
\eea
for the even modes, and 
\bea
\bar\alpha_s^m&=&\frac{1}{2\pi}\int_{-\pi}^{\pi}\mathrm{d}\theta\,\,\hat{\xi}_{(m)}^A\left[ -\left(1+\frac{s}{m+1}\right)\varepsilon_{AB}\pe_{(s+1)}^B+
 \left(1-\frac{s}{m+1}\right)R^/_{AB}\pe_{(s-1)}^B \right]\,,\\
\bar\beta_s^m&=& \frac{1}{2\pi}\int_{-\pi}^{\pi}\mathrm{d}\theta\,\,\hat{\xi}_{(m)}^A\left[ -\left(1+\frac{s}{m+1}\right)\pe^{(s+1)}_A+
 \left(1-\frac{s}{m+1}\right)R^-_{AB}\pe_{(s-1)}^B \right]\,,
\eea
for the odd modes. The spin zero modes are slightly different:
\bea
\alpha_0^m&=&\frac{1}{2\pi}\int_{-\pi}^{\pi}\mathrm{d}\theta\,\,\hat{\xi}_{(m)}^A \pe^{(1)}_A\,,\\
\bar\alpha_0^m&=&\frac{1}{2\pi}\int_{-\pi}^{\pi}\mathrm{d}\theta\,\,\varepsilon_{AB}\hat{\xi}_{(m)}^A \pe_{(1)}^B\,.
\eea

The extraction of these modes assumes $r$ is constant as a function of $\theta$. If we consider $\bm\zeta$ to be on the source plane, then the integral is taken on circles on the source plane coordinates. Alternatively, if $\bm\zeta$ is on the image plane, and the map projects back to the source, the integrals can be taken as circles on the image.

\section{Lensing roulettes in the weak field approximation}

We have shown so far that the general map which gives $\bm\xi_{(m)}$ at each order may given in terms of the spin coefficients $\alpha_s, \beta_s$ and $\bar\alpha_s, \bar\beta_s$, which are functions of radial distance in the screen space. These independent roulettes are then summed over as coefficients in a vector Fourier series to give the total map. This decomposition is completely general and will apply even if we were brave enough to use the fully non-linear GDE to calculate the maps. 

We shall now specialise to the standard weak field situation~-- where we linearise around Minkowski space~-- to illustrate how this happens.

\subsection{The weak field approximation}

We shall now linearise around Minkowski space.  We shall write our perturbations with respect to the Poisson gauge, where 
\begin{eqnarray}\label{metric}
\d  s^2 &=&
 \big[-(1 + 2\Phi )\d \eta^2 
 + (1-2 \Psi)\gamma_{i j}\d x^{i}\d x^{j}\big]\,\,.
\end{eqnarray}
The Jacobi maps in the background become
\be
{\bm{\mathcal{J}}}=\chi\bm I,~~~~~{\bm{\mathcal{K}}}=\bm I\,.
\ee
Here, $\chi$ is the radial distance.
By linearising the Jacobi map we are making an approximation which is not strictly valid in the strong field regime.
The projected part of the Riemann tensor is, to leading order in derivatives of the potential,
\be
\mathcal{R}_{AB}=\delta_{AB}-\nabla_{A}\nabla_B(\Phi+\Psi)\,.
\ee
Here, the screen space derivative acting on a scalar is 
\be
\chi\bm\nabla={\hat{\bm e}_\vartheta}\partial_\vartheta+\frac{\hat{\bm e}_\varphi}{\sin\vartheta}\partial_\varphi\,,
\ee
where $\vartheta,\varphi$ are coordinates about the observer. (As we are keeping only the maximum number of screen space derivatives, we can ignore any rotation coefficients which would appear in a more accurate calculation~\cite{Clarkson:2015pia}.) 
From this we can give the generic image-to-source map as (for $m\geq2$)
\bea\label{opkdvdfv}
\mathcal{M}^{{}^\searrow}_{AB_1\cdots B_m} &=& 
-\int_0^\chi\d\chi'\left(\frac{\chi}{\chi'}-1\right){\chi'}^{m+1}\nabla_A\nabla_{B_1}\cdots\nabla_{B_m}(\Phi+\Psi)\nonumber\\
&=& 
-{\chi}^{m+1}\nabla_A\nabla_{B_1}\cdots\nabla_{B_m}\int_0^\chi\d\chi'\left(\frac{\chi}{\chi'}-1\right)(\Phi+\Psi)\,,
\eea
while the source-to-image map is
\bea
\mathcal{M}^{{}^\nwarrow}_{AB_1\cdots B_m} &=&
-\int_0^\chi\d\chi'\left(\frac{1}{\chi}-\frac{1}{\chi'}\right)\frac{{\chi'}^{m+1}}{\chi^m}\nabla_A\nabla_{B_1}\cdots\nabla_{B_m}(\Phi+\Psi)\nonumber\\
&=& 
\nabla_A\nabla_{B_1}\cdots\nabla_{B_m}\int_0^\chi\d\chi'\left(\frac{\chi}{\chi'}-1\right)(\Phi+\Psi)\,,
\eea
where we have used
\be
\chi\nabla_A\int_0^\chi\d\chi' f(\chi') = \int_0^\chi\d\chi'\chi'\nabla_A f(\chi') \,.
\ee
Strictly speaking the indices here are on the spherical screen space about the observer, which is why they do not commute with the integrals. To convert to a flat sky approximation with Cartesian coordinates we replace $\nabla_A\mapsto\partial_A$ on the second line in each expression~-- i.e., with the derivatives outside the integral. (This is because swapping $\nabla$ and the integrals involves commuting a tetrad vector with the integral. Then, in coordinates, $\mathcal M_{ij\cdots}=e_i^Ae_j^B\cdots \mathcal M_{AB\cdots}$, and $\partial_i=e^A_i\nabla_A$, etc.) 
The Jacobi map arrises when $m=1$ and is normally defined as the map from observer to source. Using~\eqref{opkdvdfv} to find the perturbed part of the Jacobi map,  gives:
\be
\mathcal{J}_{AB}=
\chi\mathcal{A}_{AB}=\chi\delta_{AB}-{\chi}^{2}\nabla_A\nabla_{B}\int_0^\chi\d\chi'\left(\frac{\chi}{\chi'}-1\right)(\Phi+\Psi)\,,
\ee
where we have defined the conventional amplification matrix $\bm{\mathcal{A}}$ in the middle. Recognising that $\chi\nabla_A$ is a derivative on the observers celestial sphere, we define the lensing potential as
\be
\psi=\int_0^\chi\d\chi'\left(\frac{\chi-\chi'}{\chi\chi'}\right)(\Phi+\Psi)\,,
\ee 
giving the amplification matrix as
\be
\mathcal{A}_{AB}=\delta_{AB}-\chi^2\nabla_A\nabla_B\psi\,.
\ee
(Note that many authors employ a flat sky approximation replacing $\nabla_A\mapsto\partial_A$ and letting $A$ be a coordinate index before commuting the derivative with the integral~-- this gives a different form for the lensing potential.  )
We therefore have 
\bea
-\mathcal{M}^{{}^\nwarrow}_{AB_1\cdots B_m} & =&  \chi^{-(m+1)}\mathcal{M}^{{}^\searrow}_{AB_1\cdots B_m} \nonumber\\&
=& -\chi\nabla_A\nabla_{B_1}\cdots\nabla_{B_m}\psi \nonumber\\&
= &\chi^{-1}\nabla_A\nabla_{B_1}\cdots\nabla_{B_{m-2}}\mathcal{A}_{B_{m-1}B_{m}}\,.\label{kdjsncdkjncskn}
\eea
Given this, let us define the dimensionless amplification tensors as an extension of the amplification matrix:
\bea
\mathcal{A}_{AB_1\cdots B_m} &=& -\chi^{m+1}\nabla_A\nabla_{B_1}\cdots\nabla_{B_m}\psi \nonumber\\&
=&\chi^{-1}\mathcal{M}^{{}^\searrow}_{AB_1\cdots B_m}\nonumber\\&
 =& -\chi^m \mathcal{M}^{{}^\nwarrow}_{AB_1\cdots B_m}\,.\label{dksjncskjcdsdn}
\eea
Constructing the normal modes of this dimensionless tensor will then trivially give both the image-to-source and source-to-image maps. 
The fact that this dimensionless map is just angular derivatives of the amplification matrix recovers the usual approach to higher-order lensing such as flexion, as a Taylor expansion of the amplification matrix~-- but this is only valid in the approximations used here.

\subsection{Circularly symmetric lens in the thin lens approximation}

\begin{figure}[htbp]
\begin{center}
\includegraphics[width=0.6\textwidth]{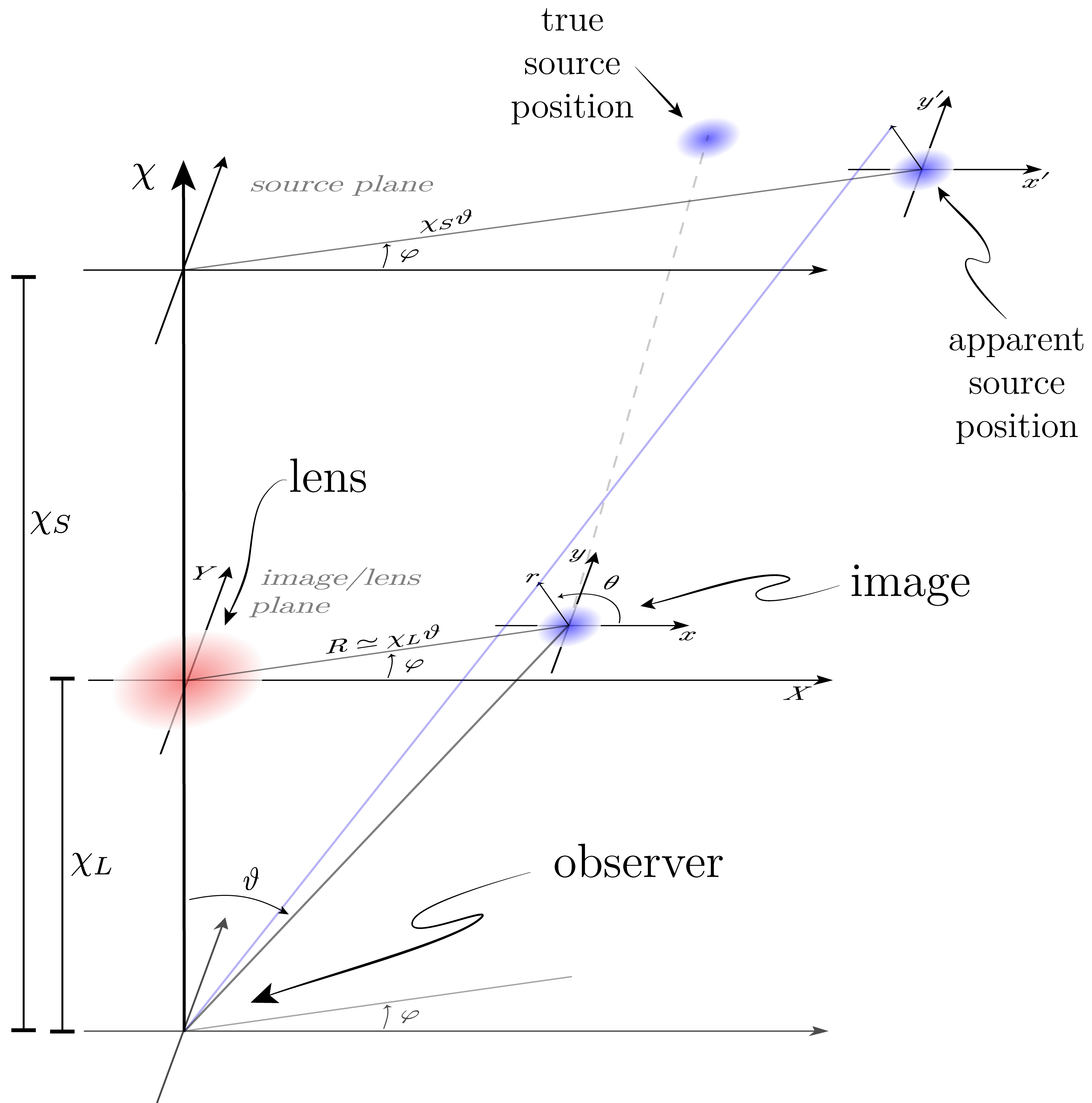}
\caption{The coordinates used in the thin lens approximation.}
\label{kdjkshbvsiuhcdjscnalc}
\end{center}
\end{figure}

The simplest case to consider is that of a rotationally symmetric lens such as a point source or an isothermal sphere. Our aim is to give the amplitudes of the normal modes $\alpha_s^m$ and $\beta_s^m$ as functions of distance from the lens. We shall do these for the dimensionless map $\mathcal{A}_{AB_1\cdots B_m}$ since the source-to-observer and observer-to-source modes being trivially found from~\eqref{dksjncskjcdsdn}. Anticipating the thin lens approximation,  we shall use Cartesian $X,Y$ coordinates centred on the lens in the lens plane, with the distance from the lens centre $R=\sqrt{X^2+Y^2}\simeq\chi_L\vartheta$, where $\vartheta$ is the angle at the observer from the centre of the lens. For a circularly symmetric lens we have the lensing potential $\psi=\psi(R)$. We denote $\psi'=\partial_R\psi(R)$. (For a more general lens in the flat sky approximation, we can just replace $\psi(R)\to\psi(X,Y;\chi)$ and $\chi_L\to\chi$ in these expressions below~-- $X,Y$ are then coordinates at distance $\chi$ from the observer.) We illustrate the coordinates we use in Fig.~\ref{kdjkshbvsiuhcdjscnalc}.


To construct $\alpha_s^m$ and $\beta_s^m$ we require
\bea\label{siudhcidhcusdhi}
\hat{\xi}_A^{{(m)}}&=&\mathcal{A}_{AB_1\cdots B_m}\hat\zeta^{B_1}\cdots\hat\zeta^{B_m}\nonumber\\&=&-\chi_L^{m+1}\hat\zeta^{B_1}\cdots\hat\zeta^{B_m}\nabla_A\nabla_{B_1}\cdots\nabla_{B_m}\psi(R)\nonumber\\
&=&-\chi_L^{m+1}\sum_{k=0}^m {m\choose k} \cos^{m-k}\theta\sin^k\theta\,\, \partial_A\partial_{X}^{m-k}\partial_Y^k\psi(R)\bigg|_\text{source position}\,.
\eea
Then we have, for $s>0$,
\bea
\alpha^m_s&=&\frac{1}{\pi}\int_{-\pi}^\pi\d\theta \left(\hat{\xi}_X^{(m)}\cos\theta +\hat{\xi}_Y^{(m)}\sin\theta \right)\cos s\theta\nonumber\\
&=&-\chi_L^{m+1}\sum_{k=0}^m {m\choose k} \bigg[
\partial_{X}^{m-k+1}\partial_Y^k\psi(R) 
\left(\frac{1}{\pi}\int_{-\pi}^\pi\d\theta \sin^k\theta\cos^{m-k+1}\theta\cos s\theta\right)
\nonumber\\&&+
\partial_{X}^{m-k}\partial_Y^{k+1}\psi(R)
\left(\frac{1}{\pi}\int_{-\pi}^\pi\d\theta \sin^{k+1}\theta\cos^{m-k}\theta\cos s\theta\right)
\bigg]\,.
\eea
The integrals do not evaluate trivially so we define the symbols
\bea
\C{k}{m}{s}&=&\frac{1}{\pi}\int_{-\pi}^\pi\d\theta\sin^{k}\theta\,\cos^{m-k+1}\theta\,\cos s\,\theta\,,
\eea
and derive the formula for evaluating them in Appendix~\ref{sdnckjsndcsdn}. We have
\bea
\alpha^m_s&=&-\chi_L^{m+1}\sum_{k=0}^m {m\choose k} \bigg[
\C{k}{m}{s} \partial_X
+
\C{k+1}{m}{s}\partial_Y
\bigg]\partial_{X}^{m-k}\partial_Y^k\psi(R)\,.\label{djksnckjdncsk}
\eea
Similarly, we find
\bea
\beta^m_s&=&-\chi_L^{m+1}\sum_{k=0}^m {m\choose k} \bigg[
\S{k}{m}{s}\partial_{X} 
+
\S{k+1}{m}{s}\partial_Y
\bigg]\partial_{X}^{m-k}\partial_Y^k\psi(R)\,,\label{djknjkdnvkdjfv}
\eea
where 
\bea
\S{k}{m}{s}&=&\frac{1}{\pi}\int_{-\pi}^\pi\d\theta\sin^{k}\theta\,\cos^{m-k+1}\theta\,\sin s\,\theta\,.
\eea

Eqs.~\eqref{djksnckjdncsk} and \eqref{djknjkdnvkdjfv} are the general formula for the amplitudes of the lensing maps at any order or spin, under the assumption of a weak gravitational field and a thin lens with circular symmetry. The odd modes are of course zero since the amplification tensors are symmetric, and the odd modes arise from one anti-symmetric interchange on the indices. It takes a bit of work to show that they are zero from \eqref{siudhcidhcusdhi}, as the $k$'th term cancels with part of the $k-1$'th and $k+1$'th terms, on using the identities \eqref{dcksbcskbcsbc}.

We tabulate the first few of these moments below. To simplify this we define the magnitude of a mode as
\be
\mu^m_s=\sqrt{(\alpha^m_s)^2+(\beta^m_s)^2}
\ee
which for a circularly symmetric lens only depends on $R$ not on $X$ or $Y$ individually. The square-root when defining $\mu^m_s$ below is taken symbolically with the convention that the sign in front of the $\psi'$ term is positive. This may not give a positive definite magnitude in all cases. 

 First, $m=1$ corresponding to normal linear lensing:
\bea
\text{spin-0:}~~~~\mu^1_0&=&\frac{\chi_L^2}{2R}(\psi'+R\psi'') = \kappa\,,\\
\alpha^1_0&=&-\mu^1_0\,,~~~~
\beta^1_0=0\,,\\
\text{spin-2:}~~~~\mu^1_2&=& \frac{\chi_L^2}{2R}(\psi'-R\psi'')\,,\\
\alpha^1_2&=& \frac{X^2-Y^2}{R^2}\mu^1_2 = -\gamma_1\,,~~~~
\beta^1_2= \frac{2XY}{R^2}\mu^1_2 = -\gamma_2\,.
\eea
We have given the relation with the convergence $\kappa$ and shear $\gamma_1,\gamma_2$ as usually defined~\cite{Lasky:2009ca}. 
Note that the spin-1 mode is zero in the approximation used in this paper so we do not list it. 
Now for $m=2$ corresponds to weak lensing flexion:
\bea
\text{spin-1:}~~~~\mu^2_1&=&\frac{3\chi_L^3}{4R^2}(\psi'-R\psi''-R^2\psi''') = -\frac{3}{2}\mathcal{F}\,,\\
\alpha^2_1&=&\frac{X}{R}\mu^2_1\,,~~~~
\beta^2_1= \frac{Y}{R}\mu^2_1\,,\\
\text{spin-3:}~~~~\mu^2_3&=&\frac{\chi_L^3}{4R^2}(3\psi'-3R\psi''+R^2\psi''')=\mathcal{G}\,,\\
\alpha^2_3&=&\frac{X(-X^2+3Y^2)}{R^3}\mu^2_3\,,~~~~
\beta^2_3=\frac{Y(-3X^2+Y^2)}{R^3}\mu^2_3\,.
\eea
Again we list only the non-zero spin modes. Here we have made the correspondence with the commonly used $\mathcal{F}$ and $\mathcal{G}$ flexion amplitudes~\cite{Lasky:2009ca}.

Now for the first new modes~-- $m=3$ has
\bea
\text{spin-0:}~~~~\mu^3_0&=&\frac{3\chi_L^4}{8R^3}(\psi'-R\psi''+2R^2\psi'''+R^3\psi'''')\,\\
\alpha^3_0&=&-\mu^3_0\,~~~~
\beta^3_0=0\,,\\
\text{spin-2:}~~~~\mu^3_2&=& \frac{\chi_L^4}{2R^3}(3\psi'-3R\psi''+R^3\psi'''')\,,\\
\alpha^3_2&=& \frac{-X^2+Y^2}{R^2}\mu^3_2\,,~~~~
\beta^3_2= -\frac{2XY}{R^2}\mu^3_2\,,\\
\text{spin-4:}~~~~
\mu^3_4&=&\frac{\chi_L^4}{8R^3}
(15\psi'-15R\psi''+6R^2\psi'''-R^3\psi'''')\,,\\
\alpha^3_4&=&\frac{X^4-6X^2Y^2+Y^4}{R^4}
\mu^3_4\,,~~~~
\beta^3_4=\frac{4XY(X^2-Y^2)}{R^4}
\mu^3_4\,.
\eea
$m=4$:
\bea
\text{spin-1:}~~~~\mu^4_1&=&\frac{5\chi_L^5 }{8R^4}(3\psi'-3R\psi''+3R^2\psi'''-2R^3\psi''''-R^4\psi''''')\,,\\
\alpha^4_1&=&\frac{X}{R}\mu^4_1\,,~~~~
\beta^4_1=\frac{Y}{R}\mu^4_1\,,\\
\text{spin-3:}~~~~\mu^4_3&=&\frac{5\chi_L^5}{16R^4}(15\psi'-15R\psi''+3R^2\psi'''+2R^3\psi''''-R^4\psi''''')\,,\\
\alpha^4_3&=&\frac{X(X^2-3Y^2)}{R^3}\mu^4_3\,,~~~~
\beta^4_3=\frac{Y(3X^2-Y^2)}{R^3}\mu^4_3\,,\\
\text{spin-5:}~~~~\mu^4_5&=&\frac{\chi_L^5}{16R^4}(105\psi'-105R\psi''+45R^2\psi'''-10R^3\psi''''+R^4\psi''''')\,,\\
\alpha^4_5&=& \frac{X(-X^4+10X^2Y^2-5Y^4)}{R^5}\mu^4_5\,,~~~~
\beta^4_5=\frac{Y(-5X^4+10X^2Y^2-Y^4)}{R^5}\mu^4_5\,.
\eea
And so on.

\subsection{Mass distribution and reconstruction in the thin lens approximation}

We can relate these to the lens mass distribution, following~\cite{Lasky:2009ca}. Ignoring peculiar velocity terms we have the Poisson equation $\nabla^2\Phi=\frac{1}{2}\rho$ (in units $8\pi G=1=c$) giving
\bea
\psi(\chi_S)=\int_0^{\chi_s}\d\chi'\left(\frac{\chi_S-\chi'}{\chi_S\chi'}\right)\nabla^{-2}\rho(\chi',R)\simeq\left(\frac{\chi_S-\chi_L}{\chi_S\chi_L}\right)\nabla^{-2}\int_0^{\chi_S}\d\chi'\rho(\chi',R)\,,
\eea
where $\nabla^{-2}$ is the inverse 3D Laplacian. In the thin lens approximation we can approximate the integral by the second step.  Writing 
\be
\Sigma(R)=\int_0^{\chi_S}\rho(\chi,R),~~~\text{and}~~~\Sigma_\text{cr}=\frac{2\chi_S}{\chi_L(\chi_S-\chi_L)}
\ee
as the projected surface mass density and the critical surface density respectively we have
\be
\psi(R)=\frac{2}{\chi_L^2}\int_0^R\frac{\d R_2}{R_2}\int_0^{R_2}\d R_1 R_1\frac{\Sigma(R_1)}{\Sigma_\text{cr}}\,.
\ee
The projected mass distribution is given by 
\be
M(R)=2\pi\int_0^R\d R_1 R_1\Sigma(R_1)\,,
\ee
so that the lensing potential is a first integral of the mass:
\be
\psi(R)=\frac{1}{\pi\chi_L^2\Sigma_\text{cr}}\int_0^R\d R_2\frac{M(R_2)}{R_2}\,.
\ee
In terms of the mass, the magnitude of each mode is
\bea
\mu^1_0&=&\frac{1}{2\pi R\Sigma_\text{cr}}\left[ M' \right]\,,\\
\mu^1_2&=&\frac{1}{2\pi R\Sigma_\text{cr}}\left[ 2\frac{M}{R}-M' \right]\,,\\
\mu^2_1&=&\frac{3}{4}\frac{\chi_L}{\pi R^2\Sigma_\text{cr}}\left[M'-RM''\right]\,,\\
\mu^2_3&=&\frac{\chi_L}{2\pi R^2\Sigma_\text{cr}}\left[\frac{4M}{R}-\frac{5}{2}M'+\frac{1}{2}RM''\right]\,,\\
\mu^3_0 &=& \frac{3}{8}\frac{\chi_L^2}{\pi R^3\Sigma_\text{cr}}\left[M'-RM''+R^2M'''\right]\,,\\
\mu^3_2 &=&\frac{\chi_L^2}{2\pi R^3\Sigma_\text{cr}}\left[ 3M'-3RM''+R^2M'''\right]\,,\\
\mu^3_4 &=&\frac{\chi_L^2}{2\pi R^3\Sigma_\text{cr}}\left[ \frac{12M}{R}-\frac{33}{4}M'+\frac{9}{4}RM''-\frac{1}{4}R^2M'''\right]\,,\\
\mu^4_1 &=& \frac{5}{8}\frac{\chi_L^3}{\pi R^4\Sigma_\text{cr}}\left[3M'-3RM''+2R^2M'''-R^3M''''\right]\,\\
\mu^4_3 &=&\frac{5}{16}\frac{\chi_L^3}{\pi R^4\Sigma_\text{cr}}\left[ 15M'-15RM''+6R^2M'''-R^3M''''\right]\,,\\
\mu^4_5 &=&\frac{\chi_L^3}{\pi R^4\Sigma_\text{cr}}\left[ \frac{24M}{R}+\frac{1}{16}(-279M'+87RM''-14R^2M'''+R^3M'''')\right]\,. 
\eea
And so on~-- we list 2 new types of modes at order 3 and 4 to observe the patterns that emerge at higher order. 

If we consider the lhs of these equations as observables, we can invert this system to give the derivatives of the lens mass distribution directly as linear combinations of them, together with a set of constraints between the $\mu^m_s$'s. The mass and its derivatives are given by (the terms are dimensionless)
\bea
\frac{M}{2\pi R^2\Sigma_\text{cr}} &=& \frac{1}{2}\mu^1_2+\frac{1}{2}\mu^1_0\,,\\
\frac{M'}{2\pi R\Sigma_\text{cr}} &=&  \mu^1_0\,,\\
\frac{M''}{2\pi \Sigma_\text{cr}} &=& -\frac{2}{3}\left(\frac{R}{\chi_L}\right)\mu^2_1+\mu^1_0\,,\\
\frac{RM'''}{2\pi \Sigma_\text{cr}} &=& \frac{4}{3}\left(\frac{R}{\chi_L}\right)^2\mu^3_0-\frac{2}{3}\left(\frac{R}{\chi_L}\right)\mu^2_1\,,\\
\frac{R^2M''''}{2\pi \Sigma_\text{cr}} &=& \frac{8}{3}\left(\frac{R}{\chi_L}\right)^2\mu^3_0+\frac{2}{3}\left(\frac{R}{\chi_L}\right)\mu^2_1-\frac{4}{5}\left(\frac{R}{\chi_L}\right)^3\mu^4_1\,,\\
\frac{R^3M'''''}{2\pi \Sigma_\text{cr}} &=& -\frac{4}{3}\left(\frac{R}{\chi_L}\right)^2\mu^3_0
-\frac{4}{3}\left(\frac{R}{\chi_L}\right)\mu^2_1-\frac{8}{5}\left(\frac{R}{\chi_L}\right)^3\mu^4_1+\frac{8}{5}\left(\frac{R}{\chi_L}\right)^4\mu^5_0\,.
\eea
Higher mass derivatives can in principle be reconstructed from the higher order roulettes. These relations can be viewed as a set of differential consistency conditions that the $\mu^m_s$'s have to satisfy for a circular lens, since each equation is the derivative of the one above.  The higher spin moments are related to the lower spin ones via the algebraic consistency conditions:
\bea
\mu^2_3&=&2\left(\frac{\chi_L}{R}\right)\mu^1_2-\frac{1}{3}\mu^2_1\,,\\
\mu^3_2&=&\frac{4}{3}\mu^3_0+\frac{4}{3}\left(\frac{\chi_L}{R}\right)\mu^2_1\,,\\
\mu^3_4&=&-\frac{1}{3}\mu^3_0-\frac{4}{3}\left(\frac{\chi_L}{R}\right)\mu^2_1+6\left(\frac{\chi_L}{R}\right)^2\mu^1_2\,,\\
\mu^4_3&=&\frac{1}{2}\mu^4_1+\frac{10}{3}\left(\frac{\chi_L}{R}\right)\mu^3_0+\frac{10}{3}\left(\frac{\chi_L}{R}\right)^2\mu^2_1\,,\\
\mu^4_5&=&-\frac{1}{10}\mu^4_1-2\left(\frac{\chi_L}{R}\right)\mu^3_0-6\left(\frac{\chi_L}{R}\right)^2\mu^2_1+24\left(\frac{\chi_L}{R}\right)^3\mu^1_2\,.
\eea

 \subsection{Examples of image construction for circular lenses}
 
 Here we shall give some examples to illustrate how the roulettes add together to create a complete image. We consider  examples of a source near the Einstein radius of a lens~-- where corrections to the amplification matrix are of order unity, and we enter the strong lensing regime. More examples can be found in Paper I.
 
 \subsubsection{Point Mass}
 
 A point mass can be characterised as $\Sigma(R)\propto\delta(R)$ implying that the projected mass $M$ is a constant (see~\cite{Virbhadra:1999nm} for a full discussion of the Schwarzschild case). For this simple case we can derive general formula for any order. The critical surface density can be recast into the Einstein radius
 \be
 \Sigma_\text{cr} = \frac{M}{\pi R_E^2} = \frac{M}{\pi \chi_L^2\vartheta_E^2}\,.
 \ee
Then the amplitude of the non-zero modes are given by
\be
\mu^m_{m+1}= m!\left(\frac{R_E}{R}\right)^2 \left(\frac{\chi_L}{R}\right)^{m-1}\,.
\ee
If we place our source on the $X$ axis then we have $\beta^m_s=0$ and $\alpha^m_s=(-1)^s\mu^m_s$, simplifying things considerably. Recall that in this formalism these are considered constants over the source/image and so $R$ will be the location of the centre of the source (say) and does not vary over it.

What do these roulettes do to a source? 
Consider the image-to-source case. If we have a point on the image plane $\bm \zeta = \zeta \hat{\bm\zeta} = (x,y)/\chi_L$, where $\zeta=r/\chi_L$ is the  radial coordinate in the image plane (or the angular distance measured by the observer) in coordinates centred at the image, this is mapped to a location on the source plane given by 
\bea
(\searrow)~~~~~~~~~~\xi^\text{(source)}_A &=& \chi_S\zeta_A + \chi_S\sum_{m=1}^\infty \frac{\zeta^m}{m!} \mathcal{A}_{AB_1\cdots B_m}\hat\zeta^{B_1}\cdots\hat\zeta^{B_m}\,.
\eea
which becomes in coordinates
\bea\label{dskjnckjdscdnsdjknvnsd}
(\searrow)~~~~~~~~~~\left(\frac{\chi_L}{\chi_S}\right){x'\choose y'} = r{\cos\theta\choose\sin\theta}
-\frac{R_E^2}{R}\sum_{m=1}^\infty (-1)^{m} \left(\frac{r}{R}\right)^{m}
{\cos m\theta\choose-\sin m\theta}\,.
\eea
If we consider this as a set of parametric equations for $x',y'$, lines of $r=\text{const.}$ will be shapes of sources which create circular images. More usefully, if we consider an intensity profile which is a function of ${x'}^2+{y'}^2$ then the level surfaces of this function in $(x,y)$ image plane coordinates will be images formed corresponding to circles in the source plane.

These series will only absolutely converge for all angles for $r<R$, which places a limit on the size of the source one can usefully consider. This is only important for a large source observed close to the Einstein radius, but it also means that secondary images cannot be reconstructed using the same series as the primary image; a second series with different $X,Y$ coordinates at the centre of each image can be used for multiple images. The series can be summed however, and analytically continued outside of this region by writing $r^m\cos m\theta=\Re{(r^me^{\i m\theta})}$, etc.,  and using the geometric series. This gives:
\be
(\searrow)~~~~~~~~~~\left(\frac{\chi_L}{\chi_S}\right){x'\choose y'} = r{\cos\theta\choose\sin\theta}
+\frac{R_E^2 r}{r^2+{R^2}+2rR\cos\theta}\left(
\begin{array}{c}
{\displaystyle\frac{r}{R}+\cos\theta}
\\ 
-{\sin\theta}
\end{array}
\right)
\ee
This is the usual equation for lensing from a point mass in slightly unfamiliar form, with the centre of the source shifted by $(\chi_S/\chi_L)(R_E^2/R)$. 
 The key difference from the usual form of this equation is that the formalism here uses a sort of Born-like approximation: the light on the centre of the source travels on a straight line, so given an image position,  the source position is given in the `wrong place', but with the correct shape.

Consider a large source with a Gaussian intensity profile $e^{-({x'}^2+{y'}^2)/2\sigma^2}$ with $\sigma=1/6 (\chi_L/\chi_S) R_E$, located near the Einstein radius, with $R=1.3R_E$ in the lens plane
(corresponding to a true unlensed position of $\simeq(\chi_S/\chi_L) 0.53 R_E$ from the centre of the lens).  
 This is something which should normally be considered in the strong lensing regime, but we can accurately  recover the principal image using enough roulettes. (The secondary image is recovered from the analytic continuation of the full sum of roulettes.) Given that the series is divergent for $r\geq R$, we expect (and find) problems at this radius. Even though this is a fairly trivial example, the series tells us a lot about the roulettes which go into making the whole image.

\begin{figure}[htbp]
\begin{center}
\includegraphics[width=\textwidth]{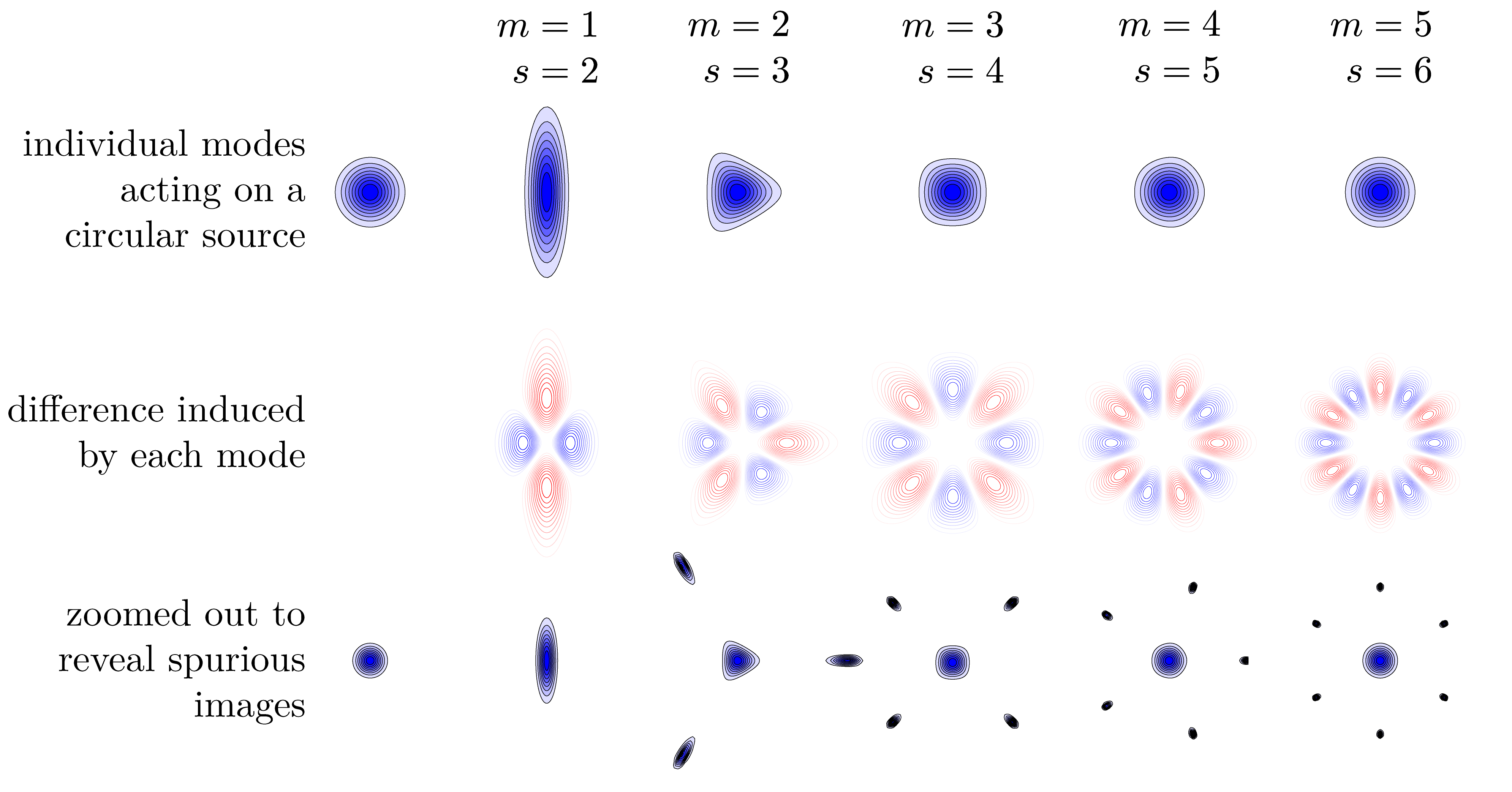}
\caption{A circular source located at apparent location $R=1.3R_E$ from a point mass, with width $\sigma=1/6 (\chi_L/\chi_S) R_E$ (corresponds to a full width at half maximum of about $0.4 (\chi_L/\chi_S) R_E$). Here we consider the effect of each mode individually. In the second row red is positive and blue negative (the relative amplitudes between plots is not fixed; the number of contours is fixed instead). }
\label{dskjdkjsvnjknv}
\end{center}
\end{figure}

Consider Fig.~\ref{dskjdkjsvnjknv}. This takes each mode individually acting on the source. In the top row we see the familiar shearing into an ellipse, and flexion distorting into a triangular shape. Higher roulettes distort into bulging squares, pentagons, hexagons, etc., with the symmetry of the shape determined by the spin of the mode. The spin is clearly seen in the second row where the initial image is subtracted off. 
In the final row we have zoomed out from the central image to reveal $s$ spurious images originating outside the radius of convergence. These images occur at points where ${x'}^2+{y'}^2=0$, which for the $m$'th roulette is when
\be
\frac{r}{R_E}=\left(\frac{R}{R_E}\right)^{(m+1)/(m-1)}~~~\text{and}~~~\theta=\frac{m-2k}{m+1}\pi,~~~k\in \mathbbm{Z}\,,
\ee
and in the limit of large $m$ these approach the radius of convergence, forming a ring of images around the source which must be removed. For extended sources, particularly near the Einstein radius, these extra images blend into the principal image. 

\begin{figure}[htbp]
\begin{center}
\includegraphics[width=\textwidth]{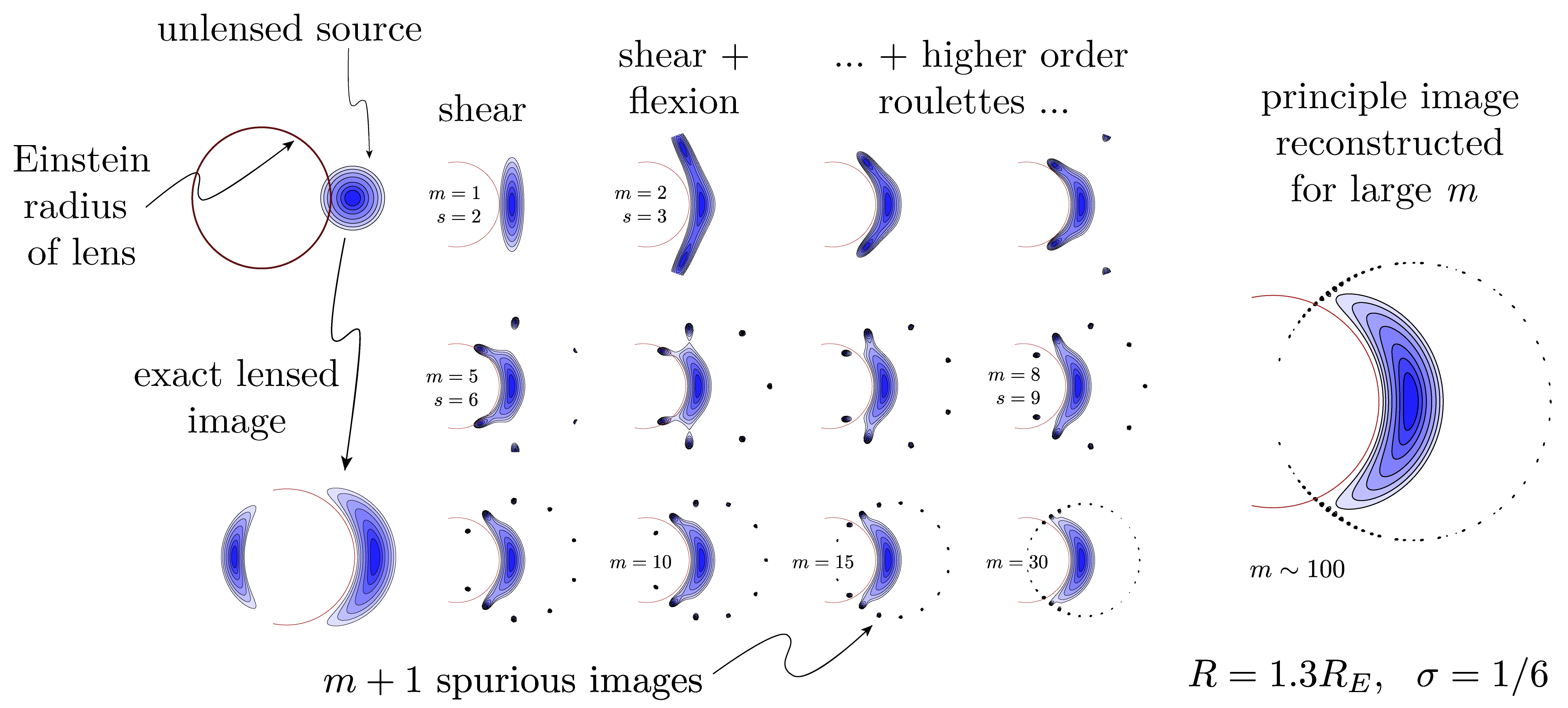}
\caption{The same lens-source configuration as in Fig.~\ref{dskjdkjsvnjknv}. Here we consider the effect of summing up to a given mode indicated, so show how the series converges to give the full principal image. Top left we show the unlensed source, which has a strongly lensed image shown bottom left. In the middle panel of 12 figures we add more roulettes to each image (from left to right, then top to bottom), with $m$ indicating the number of roulettes added. On the right we show a fully formed image surrounded by a ring of $m+1$ spurious images. }
\label{sdkjcndjal}
\end{center}
\end{figure}

Now let us consider how these images `sum' to give a complete lensed image, illustrated in Fig.~\ref{sdkjcndjal}, using the image-to-source map $(\searrow)$. 
We consider the same lens-source set-up, but now add progressively more roulettes to the image. In this example we see that the shear and flexion give very poor approximations to the correct image. As we add more roulettes, the image takes shape by around $m\sim10$, but has problems around the edge where the image blends into the spurious images. However, for $m$ sufficiently large the principal image and the spurious images split apart, and we see the true image appear fully resolved. The secondary image is outside the radius of convergence and is not recovered. This can be found instead by choosing $X,Y$ at the centre of that image instead.

An alternative way to look at the convergence of the series is to consider the partial sum roulettes in \eqref{dskjnckjdscdnsdjknvnsd} as parametric equations for $(x',y')$ for fixed $r$ and parameter $\theta\in[0,2\pi)$. 
\begin{figure}[htbp]
\begin{center}
\includegraphics[width=\textwidth]{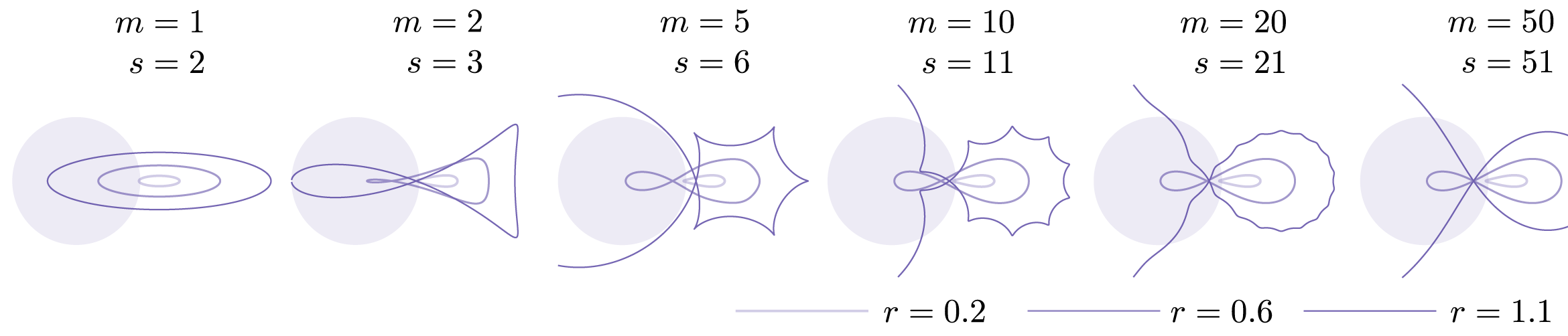}
\caption{ Shapes in the source plane that get distorted into a circle of radius shown (with $(\chi_L/\chi_S) R_E=1$). It's not a  physically relevant scenario, but we see the equivalent of the spurious images are the cusps in the curves with large radius. Adding more maps cancels out these cusps order by order.  }
\label{sdkjcnsdcjknadsdjal}
\end{center}
\end{figure}
Such curves show the curves in the source plane which would be distorted into a circle in the image plane.

 \subsubsection{Singular Isothermal Sphere}
 
For an isothermal sphere, we have,~\cite{Lasky:2009ca},
\be
\Sigma(R)=\frac{R_E}{2R}\Sigma_\text{cr}~~~\Rightarrow~~~M(R)=\pi\Sigma_\text{cr}R_ER\,,~~~\psi(R)=\frac{R_ER}{\chi_L^2}\,.
\ee
The amplitudes of the roulettes cannot be found to obey a simple formula as in the point mass case, as they require the full set of the trig integrals discussed in the appendix. We tabulate the first few as
\be
\frac{2^{2m-1}R^m}{m! R_E\chi_L^{m-1}}\mu^m_s=
\left[ \begin {array}{r|cccccccccccc} 
 &s=0&1&2&3&4&5&6&7&8&9&10&11 \\ \hline 
m=1&1&0&1& & & & & & & & & \\  
2&0&3&0&3& & & & & & & & \\  
3&2&0&8&0&10& & & & & & & \\  
4&0&10&0&25&0&35& & & & & & \\  
5&12&0&30&0&84&0&126& & & & & \\
6&0&70&0&98&0&294&0&462& & & & \\  
7&100&0&224&0&336&0&1056&0&1716& & & \\  
8&0&630&0&756&0&1188&0&3861&0&6435& & \\  
9&980&0&2100&0&2640&0&4290&0&14300&0&24310& \\  
10&0&6468&0&7260&0&9438&0&15730&0&53482&0&92378
\end {array}
 \right] \,.
\ee
The important thing is the  power fall off with distance from the lens, which is one power of $R$ slower than the point mass case
\be
 \mu^m_s\sim \frac{R_E}{R}\left(\frac{\chi_L}{R}\right)^{m-1}\,,
\ee
together with the fact that for each mode the higher spin contributions are largest. By contrast with the point mass case, an SIS induces different spin modes at each order, whereas the point mass only produces modes with $s=m+1$. 
 
\begin{figure}[htbp]
\begin{center}
\includegraphics[width=\textwidth]{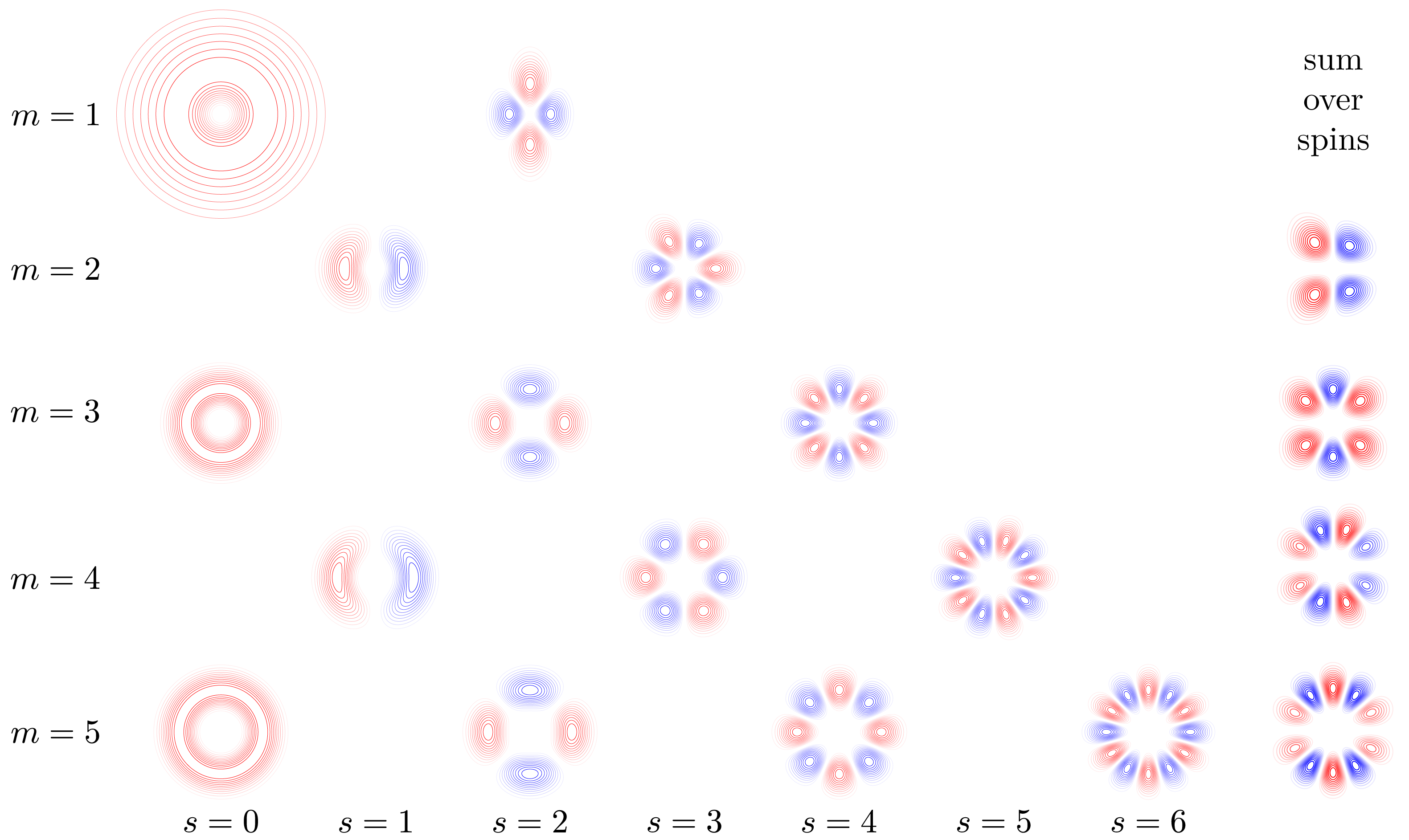}
\caption{A SIS as in Fig.~\ref{dskjdkjsvnjknv}, showing how each roulette modifies the image.}
\label{kjsdnaclkndaclk}
\end{center}
\end{figure}

In Fig.~\ref{kjsdnaclkndaclk} we consider the same physical set-up as in the previous example with a point mass lens, again using the source-to-image map. We show how each roulette modifies the circular source image, by adding a single mode and subtracting the original source. We then add up the spins for each degree to show how an image is modified order by order. 
We have spin-0 modes for odd degrees, which modify the overall size of the source. This implies that a source will be larger than the standard convergence would predict. Regarding the other modes, the roulettes with $s=m+1$ look similar to the case of a point mass lens, but slightly less distorted as expected (there is less lensing mass inside a sphere of radius $R$). Roulettes with $s<m+1$ are rather different in shape, and the changes to the image happen father from the image centre for larger $m$, and $s$ fixed. Note that these figures are not quite symmetrical, most easily seen in the $s=1$ case (the red kidney is slightly larger than the blue). 

Finally, let us consider how those modes sum to a complete image, but using the image-to-source map,  for the same physical consideration. In this case the derived equations for $(x,y)$ can be considered as parametric equations with parameter $\theta$. (The equations wind up the same, with coordinates shifted to the source plane, but with a sign change in front of the roulette sum, and we shift the lens position to $R=0.3$.) Then, lines of constant $r$ correspond to circles in the source plane and are mapped to curves in the image plane. When the image has converged we would get the same picture as using the image-to-source method previously, but the convergence approach is different so interesting to show.
\begin{figure}[htbp]
\begin{center}
\includegraphics[width=\textwidth]{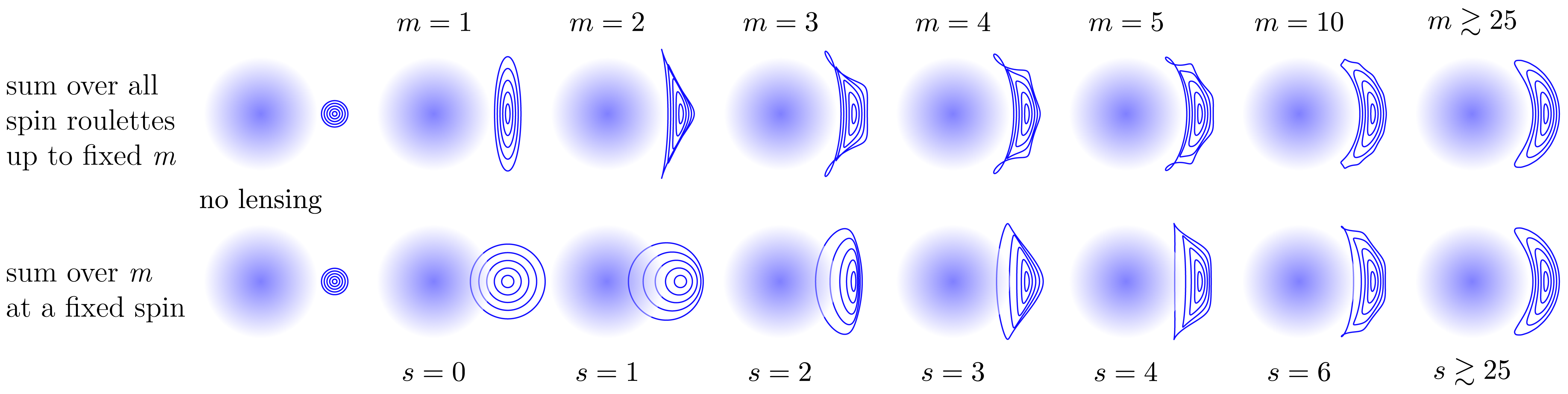}
\caption{A SIS as in Fig.~\ref{dskjdkjsvnjknv}, showing how the series of roulettes sums to give a complete image when added in different ways. The source is shown on the left and the magnified image on the right (calculated with $m=26$)~-- these are shown to scale, with the Einstein radius of the lens shown for reference. In the top  row we fix the maximum $m$ used, and add up all the possible spins. In the bottom row we add up all possible spins up to the maximum shown. }
\label{sjakdcdfhbv}
\end{center}
\end{figure}
 We consider in Fig.~\ref{sjakdcdfhbv} adding up the modes in two different ways. In the top row we add up to a fixed $m$ shown, including all the spin modes up to that order~-- this is the natural way to add up the modes as each mode contributes a factor of $R_E/R$ less. So, for $m=1$, we include the shear and convergence, and for $m=2$ we add in the flexion. Near the centre of the image this is reasonably accurate, but far from it it is not. For large $m$ we recover the exact image. It is also instructive to add in each spin type one at a time~-- we show this in the second row. We fix $m=26$, and add up all the spin-0 modes, then the spin-1 modes and so on, until the image is constructed. Note that the wobbles and cusps on the outer curves for small $m$ are the equivalent of the spurious images found in the image-to-source case above, and curves of larger radius display extravagant loops.  Even with $m\sim25$, larger circles than shown have still not converged. The image shown here corresponds to that found by using the exact solution directly.

\section{Summary and conclusions}

We have presented a new extension of the usual weak lensing formalism which can be accurately used in the strong lensing regime to reconstruct the  image(s) of the source. By expanding the general solution of the non-linear geodesic deviation equation in the screen space, one can add to the usual Jacobi map a series of lensing maps of arbitrary order, given by tensors of increasing rank. We have shown how a map at each order can be invariantly decomposed into a set of trace-free tensors representing the invariant degrees of freedom and spin modes of the map. Then we show how to reduce these trace-free tensors into distortions on the plane, which are the normal modes we call roulettes.  These extend the  convergence, shear and flexion familiar from weak lensing to arbitrary order, providing a completion of the weak lensing formalism. A complete image can be given as an appropriate two dimensional sum over these roulettes. The amplitude of each roulette can be found from integrating in a circle around a point in the image. These amplitudes are related to derivatives of the mass distribution of the lens in the weak field approximation. More generally, they are derivatives of the lensing potential perpendicular to the line of sight. We then gave a couple of examples of explicit image reconstruction for a point mass and a singular isothermal sphere, complementing those in Paper I. Although these examples are rather simple, and the exact solutions known, they are really useful to test out the method and present how an image is constructed in the strong lensing regime. An intriguing peculiarity of the roulettes is the appearance and distribution of spurious images, and understanding these in relation to the full geometry of photon surfaces might be interesting~\cite{Claudel:2000yi}.

A practical use for this formalism lies, in principle, in being able to reconstruct the mass distribution if one can extract the amplitudes and orientation of the roulettes. So, given an observation of lensed galaxies around a lens, each lensed imaged would give an estimate of the mass and its first few derivatives at the image position. From this the entire lens mass distribution could~-- in principle~-- be reconstructed. This, together with more accurate lens modelling, is left for future work.

\acknowledgments 

I would like to thank David Bacon and Julien Larena for discussions. This work is funded by the National Research Foundation (South Africa).

\appendix

\section{Notes on the trig integrals}
\label{sdnckjsndcsdn}

Here we collect together some notes on the trig integrals appearing in the spin decomposition of the maps. The key integrals are
\bea
\C{k}{m}{s}&=&\frac{1}{\pi}\int_{-\pi}^\pi\d\theta\sin^{k}\theta\,\cos^{m-k+1}\theta\,\cos s\,\theta\,,\\
\S{k}{m}{s}&=&\frac{1}{\pi}\int_{-\pi}^\pi\d\theta\sin^{k}\theta\,\cos^{m-k+1}\theta\,\sin s\,\theta\,.
\eea
These are surprisingly complicated functions of $k,m,s$.   To evaluate we use the identity
\be
\int_0^{\pi/2}\d\theta\sin^{2p-1}\theta\cos^{2p-1}\theta=\frac{\Gamma(p)\Gamma(q)}{2\Gamma(p+q)}\,,
\ee
from which it follows
\be
\frac{1}{\pi}\int_{-\pi}^{\pi}\d\theta\sin^{2p}\theta\cos^{2p}\theta=
\frac{2}{4^{p+q}}\frac{(2p)!(2q)!}{p!q!(p+q)!}\,.
\ee
For odd powers these integrals give zero. Now expand the $\cos s\theta$ and $\sin s\theta$ using
\bea
\cos s\theta&=&\sum_{k=0}^{s/2} (-1)^{k}{s\choose 2k} \cos^{n-2k}\theta\sin^{2k}\theta\,,\\
\sin s\theta&=&\sum_{k=0}^{(s-1)/2} (-1)^{k}{s\choose 2k+1} \cos^{n-2k-1}\theta\sin^{2k+1}\theta\,,
\eea
to give
\bea
\C{k}{m}{s}&=&
\left\{\begin{array}{ccc}
0&\text{if}& \text{ $k$ is odd or $s+m$ is even}\\ \\
\displaystyle\frac{1}{2^{s+m}}\sum_{j=0}^{s/2} (-1)^{j}{s\choose 2j}\frac{(2(p+j))!(2(q-j))!}{(p+j)!(q-j)!(p+q)!}\,,&\text{where}& 2p=k,~2q=s+m-k+1
\end{array}\right.
\,,\\&&\nonumber\\
\S{k}{m}{s}&=&
\left\{\begin{array}{ccc}
0&\text{if}& \text{ $k$ is even or $s+m$ is even}\\ \\
\displaystyle
\frac{1}{2^{s+m}}\sum_{j=0}^{(s-1)/2} (-1)^{j}{s\choose 2j+1}\frac{(2(p+j))!(2(q-j))!}{(p+j)!(q-j)!(p+q)!}\,,&\text{where}&2p=k+1,~2q=s+m-k
\end{array}\right.
\,.
\eea
For small values of $k,m,s$ these integrals are tabulated below. Note that $\C{0}{1}{0}=1$ and $\S{1}{2}{1}=1/4$ are the first non-zero values. 

\begin{table}[htp]
\caption{Non-zero values of $2^m\C{k}{m}{s}$. }
\begin{center}
\hspace*{-5mm}\begin{tabular}{|c|c|c|c|}
\hline
$\begin {array}{c|cc}
m=3&k=0&k=2\\ \hline s=0&6&2
\\  s=2&4&0\end {array}
$ 
& 
$\begin {array}{c|ccc} m=5&k=0&k=2&k=4\\ \hline s=0&
20&4&4\\  s=2&15&1&-1\\  s=4&6&-2&-2
\end {array}$
&
$\begin {array}{c|cccc} m=7&k=0&k=2&k=4&k=6
\\ \hline s=0&70&10&6&10\\  s=2&56&4&0&-4
\\  s=4&28&-4&-4&-4\\  s=6&8&-4&0&4
\end {array}
$ 
&
$\begin {array}{c|ccccc} m=9&k=0&k=2&k=4&k=6&k=8
\\ \hline s=0&252&28&12&12&28\\  s=2&210&
14&2&-2&-14\\  s=4&120&-8&-8&-8&-8
\\  s=6&45&-13&-3&3&13\\  s=8&10&-6&
2&2&-6\end {array}
$\\
\hline
\end{tabular}
\end{center}
\label{default}
\end{table}%

\begin{table}[htp]
\caption{Non-zero values of $2^m\S{k}{m}{s}$}
\begin{center}
\hspace*{-5mm}\begin{tabular}{|c|c|c|c|}
\hline
$\begin {array}{c|cc} m=4&k=1&k=3
\\ \hline s=1&2&2\\  s=3&3&1\end {array}
$
&
$\begin {array}{c|ccc} m=6&k=1&k=3&k=5
\\ \hline s=1&5&3&5\\  s=3&9&3&1
\\  s=5&5&-1&-3\end {array}
$
&
$\begin {array}{c|cccc} m=8&k=1&k=3&k=5&k=7\\ \hline s=1&14&6
&6&14\\  s=3&28&8&4&0\\  s=5&20&0&-4
&-8\\  s=7&7&-3&-1&5\end {array}
$
&
$\begin {array}{c|ccccc} m=10&k=1&k=
3&k=5&k=7&k=9\\ \hline s=1&42&14&10&14&42
\\  s=3&90&22&10&6&-6\\  s=5&75&5&-5
&-11&-21\\  s=7&35&-7&-5&1&19\\  s=9
&9&-5&1&3&-7\end {array}
$
\\
\hline
\end{tabular}
\end{center}
\label{default}
\end{table}%

An alternative way to calculate these is to use recurrence relations for $k, m, s$ (or just use Maple!). These are easily derivable from trig identities:
\bea
\C{k+2}{m}{s}-\C{k}{m-2}{s}+\C{k}{m}{s}&=&0\,,\\
\C{k}{m}{s-1}+\C{k}{m}{s+1}-2\C{k}{m+1}{s}&=&0
\eea
and by integrating by parts twice:
\be
(m-k+1)(m-k)\C{k+2}{m}{s}+[s^2+2k^2-(2k+1)(m+1)]\C{k}{m}{s}+k(k-1)\C{k-2}{m}{s}=0\,,
\ee
with similar identities for $\S{k}{m}{s}$.
Some identities are essential for proving that the odd modes are zero in the weak gravitational field approximation. These are found by integrating by parts once:
\bea\label{dcksbcskbcsbc}
s\S{k+1}{m}{s}&=&(k+1)\C{k}{m}{s}+(k-m)\C{k+2}{m}{s}\,,\\
s\S{0}{m}{s}&=&-(m+1)\C{1}{m}{s}\,,\\
s\S{m+1}{m}{s}&=&(m+1)\C{m}{m}{s}\,.
\eea
Again there are similar identities for $\C{k}{m}{s}$ which are required for $\bar\beta^m_s$.



\begin{thebibliography}{99}

\bibitem{Clarkson:2016zzi} 
  C.~Clarkson,
  Class.\ Quant.\ Grav.\  {\bf 33}, no. 16, 16LT01 (2016)
  [arXiv:1603.04698 [astro-ph.CO]].
  
\bibitem{Goldberg:2004hh} 
  D.~M.~Goldberg and D.~J.~Bacon,
  Astrophys.\ J.\  {\bf 619}, 741 (2005)
  [astro-ph/0406376].

\bibitem{Bacon:2005qr} 
  D.~J.~Bacon, D.~M.~Goldberg, B.~T.~P.~Rowe and A.~N.~Taylor,
  Mon.\ Not.\ Roy.\ Astron.\ Soc.\  {\bf 365}, 414 (2006)
  [astro-ph/0504478].

\bibitem{Clarkson:2015pia} 
  C.~Clarkson,
  JCAP {\bf 1509}, no. 09, 033 (2015)
  doi:10.1088/1475-7516/2015/09/033, 10.1088/1475-7516/2015/9/033
  [arXiv:1503.08660 [gr-qc]].

\bibitem{Vines:2014oba} 
  J.~Vines,
  Gen.\ Rel.\ Grav.\  {\bf 47}, no. 5, 59 (2015)
  doi:10.1007/s10714-015-1901-9
  [arXiv:1407.6992 [gr-qc]].

\bibitem{pawley}
M.~G.~Pawley,
Journal of the Franklin Institute, {\bf 307} 2 155 (1979)


\bibitem{Castro:2005bg} 
  P.~G.~Castro, A.~F.~Heavens and T.~D.~Kitching,
  Phys.\ Rev.\ D {\bf 72}, 023516 (2005)
  doi:10.1103/PhysRevD.72.023516
  [astro-ph/0503479].

\bibitem{Lasky:2009ca} 
  P.~Lasky and C.~Fluke,
  Mon.\ Not.\ Roy.\ Astron.\ Soc.\  {\bf 396}, 2257 (2009)
  [arXiv:0904.1440 [astro-ph.CO]].

\bibitem{Virbhadra:1999nm} 
  K.~S.~Virbhadra and G.~F.~R.~Ellis,
  Phys.\ Rev.\ D {\bf 62}, 084003 (2000)
  doi:10.1103/PhysRevD.62.084003
  [astro-ph/9904193].

\bibitem{Claudel:2000yi} 
  C.~M.~Claudel, K.~S.~Virbhadra and G.~F.~R.~Ellis,
  J.\ Math.\ Phys.\  {\bf 42}, 818 (2001)
  doi:10.1063/1.1308507
  [gr-qc/0005050].
 
\end{thebibliography}
\end{document}